\documentclass[acmsmall,screen]{acmart}

\usepackage{microtype}

\usepackage{pifont}
\usepackage{amsthm}
\usepackage{tikz}

\usepackage{booktabs}
\usepackage{makecell}
\usepackage{hyperref}

\usetikzlibrary{positioning}

\definecolor{mygreen}{rgb}{1, 0, 0.6}
\newcommand{\code}[1]{{\ttfamily \small #1}}

\newcommand*\rot{\rotatebox{90}}

\newcounter{suggestion}[section]

\newcounter{insight}[section]

\usepackage{tcolorbox}
\newenvironment{answerbox}{
\begin{tcolorbox}[colback=blue!5!white,colframe=blue!5!white,arc=0mm,left=1.5mm,right=1.5mm,top=0mm,bottom=0mm]
}
{
\end{tcolorbox}
}

\usepackage{tikz}
\usepackage{tikz-qtree}
\usetikzlibrary{trees} %

\usepackage{caption}
\usepackage{subcaption}

\usepackage{listings}
\usepackage{xcolor}

\definecolor{codegreen}{rgb}{0,0.6,0}
\definecolor{codegray}{rgb}{0.5,0.5,0.5}
\definecolor{codepurple}{rgb}{0.58,0,0.82}
\definecolor{backcolour}{rgb}{0.95,0.95,0.92}

\lstdefinestyle{mystyle}{
    backgroundcolor=\color{backcolour},
    commentstyle=\color{codegreen},
    keywordstyle=\color{magenta},
    numberstyle=\tiny\color{codegray},
    stringstyle=\color{codepurple},
    basicstyle=\ttfamily\footnotesize,
    breakatwhitespace=false,
    breaklines=true,
    captionpos=b,
    keepspaces=true,
    numbers=left,
    numbersep=5pt,
    showspaces=false,
    showstringspaces=false,
    showtabs=false,
    tabsize=2
}
\usepackage{xcolor}
\definecolor{celadon}{rgb}{0.67, 0.88, 0.69}

\def\code#1{\texttt{#1}}

\newcommand{\totComputingPlatforms}{18}

\newcommand{\totBugFixCommits}{2,140}
\newcommand{\totElementaryAnnotations}{886}
\newcommand{\totRealBugAnnotated}{223}
\newcommand{\totQuantumBugsAnnotated}{89}
\newcommand{\percQuantumBugsAnnotated}{39.9\%}
\newcommand{\totClassicalBugsAnnotated}{134}

\newcommand{\medianLoCQuantum}{8.0}
\newcommand{\medianLoCClassical}{4.0}

\newcommand{\nBugsIncOutput}{77}
\newcommand{\percIncOutput}{34.5\%}
\newcommand{\nBugsFailTest}{16}
\newcommand{\nBugsCrash}{92}
\newcommand{\percBugsCrash}{41.2\%}
\newcommand{\nBugsCrashOS}{67}
\newcommand{\nBugsCrashAPP}{25}
\newcommand{\nBugsNotFunctional}{14}
\newcommand{\nBugsFunctional}{188}
\newcommand{\nBugsMisconfiguration}{32}
\newcommand{\nBugsQuantumOneLiner}{10}
\newcommand{\nBugsClassicalOneLiner}{39}

\setcopyright{rightsretained}
\acmPrice{}
\acmDOI{10.1145/3527330}
\acmYear{2022}
\copyrightyear{2022}
\acmSubmissionID{oopsla22main-p104-p}
\acmJournal{PACMPL}
\acmVolume{6}
\acmNumber{OOPSLA1}
\acmArticle{86}
\acmMonth{4}

\begin{document}

\title[Bugs in Quantum Computing Platforms: An Empirical Study]{Bugs in Quantum Computing Platforms: An Empirical Study}

\author{Matteo Paltenghi}
\affiliation{%
   \institution{University of Stuttgart}
   \city{Stuttgart}
   \country{Germany}}
\email{mattepalte@live.it}

\author{Michael Pradel}
\affiliation{%
   \institution{University of Stuttgart}
   \city{Stuttgart}
   \country{Germany}}
\email{michael@binaervarianz.de}

\begin{abstract}
  The interest in quantum computing is growing, and with it, the importance of software platforms to develop quantum programs.
  Ensuring the correctness of such platforms is important, and it requires a thorough understanding of the bugs they typically suffer from.
  To address this need, this paper presents the first in-depth study of bugs in quantum computing platforms.
  We gather and inspect a set of \totRealBugAnnotated{} real-world bugs from \totComputingPlatforms{} open-source quantum computing platforms.
  Our study shows that a significant fraction of these bugs (\percQuantumBugsAnnotated{}) are quantum-specific, calling for dedicated approaches to prevent and find them.
  The bugs are spread across various components, but quantum-specific bugs occur particularly often in components that represent, compile, and optimize quantum programming abstractions.
  Many quantum-specific bugs manifest through unexpected outputs, rather than more obvious signs of misbehavior, such as crashes.
  Finally, we present a hierarchy of recurrent bug patterns, including ten novel, quantum-specific patterns.
  Our findings not only show the importance and prevalence bugs in quantum computing platforms, but they help developers to avoid common mistakes and tool builders to tackle the challenge of preventing, finding, and fixing these bugs.
\end{abstract}


\maketitle

\makeatletter
\lst@AddToHook{PreSet}{\normallineskiplimit=0pt}
\makeatother

\section{Introduction}

Quantum computing has been making immense progress recently.
Over the past decades, the field has evolved from an idea that seemed a long way off to a domain with huge investments from both public and private players~\cite{OverviewQuantumInitiatives2021}.
Quantum algorithms~\cite{shorPolynomialTimeAlgorithmsPrime1999, groverFastQuantumMechanical1996, harrowQuantumAlgorithmSolving2009,farhiQuantumApproximateOptimization2014} are on their way to becoming, in selected domains, a serious competitor for classical computing.

Executing a quantum algorithm on a quantum computer or a simulator requires a complex software stack.
We call this software stack a \emph{quantum computing platform}.
Such a platform encompasses a quantum programming language, a compiler, and an execution environment that supports running quantum programs.
Various efforts currently compete for building quantum computing platforms, e.g., Qiskit by IBM, Circ by Google, and Q\# by Microsoft.
Given the key role of quantum computing platforms in this growing domain, ensuring the correctness of these platforms is a high priority.
There are various approaches for preventing and finding bugs in other kinds of critical software infrastructure, e.g., for testing~\cite{compilerTestingSurvey2021} or verifying~\cite{leroyFormalVerificationRealistic2009} compilers.
We argue that quantum computing platforms play a role similar to language implementations for traditional computing, and hence deserve a similar level of attention.

An important prerequisite for preventing and detecting bugs is understanding what bugs exist in the wild.
Studies in other domains, e.g., on concurrency bugs~\cite{luLearningMistakesComprehensive2008a} or compiler bugs~\cite{DBLP:conf/issta/SunLZS16}, have proven useful to guide future work toward addressing relevant problems.
However, there currently is no detailed study of bugs in quantum programming platforms.

To fill this gap, this paper presents the first in-depth study of bugs in quantum computing platforms.
We gather a set of \totRealBugAnnotated{} real-world bugs from \totComputingPlatforms{} open-source projects, including highly popular platforms, such as Qiskit, Circ, and Q\#.
We thoroughly inspect and annotate these bugs in an iterative process, to understand
how many of them are quantum-specific,
what components of a quantum computing platform the bugs are in,
how these bugs manifest (and hence can be detected),
and how complex the corresponding bug fixes are.
Moreover, we identify a set of recurring bug patterns that highlight kinds of mistakes developers make repeatedly, even across different platforms.

\begin{figure}[t]
\begin{lstlisting}[language=Python, numbers=none]
def fold_global(circuit: QPROGRAM, stretch: float, **kwargs) -> QPROGRAM:
    """Gives a circuit by folding the global unitary of the input circuit."""
    ...
    # Fold remaining gates until the stretch is reached
    ops = list(base_circuit.all_operations())
\end{lstlisting}
\vspace{-1em}
\begin{lstlisting}[backgroundcolor=\color{pink},numbers=none]
-   num_to_fold = int(round(fractional_stretch * len(ops)))
\end{lstlisting}
\vspace{-1em}
\begin{lstlisting}[backgroundcolor=\color{celadon},numbers=none]
+   num_to_fold = int(round(fractional_stretch * len(ops) / 2))
\end{lstlisting}
\vspace{-1em}
\begin{lstlisting}[language=Python, numbers=none]
    if num_to_fold > 0:
        folded += Circuit([inverse(ops[-num_to_fold:])], [ops[-num_to_fold:]])
\end{lstlisting}
\caption{Example of a quantum bug that induces an incorrect output. \href{https://github.com/unitaryfund/mitiq/commit/5fc12d5e4688ba4006eb4a733379b61c0e9c2b45}{(UUID: 1687, from Mitiq)}}
\label{fig:bug_1687_mitiq}
\end{figure}

As an example of a quantum bug considered in our study, Figure~\ref{fig:bug_1687_mitiq} shows a problem from the Mitiq project~\cite{laroseMitiqSoftwarePackage2021}, along with the fix applied by the developers.
The function \code{fold\_global} is part of a noise mitigation technique called ``Zero-noise extrapolation'', which adds pairs of opposite operations to the program.
These additional operations are intended to make the computation longer and noisier, while preserving the overall mathematical result, with the goal to inject noise to extrapolate the result without noise.
The code adds a number of pairs of operations determined by \code{num\_to\_fold}.
In the buggy version, the code fails to consider that each folding step adds two operations, and not one, which is then fixed by dividing \code{num\_to\_fold} by two.
The bug results in adding twice the amount of operations, adding more noise than intended.
The example is representative for several commonly observed characteristic of bugs in quantum computing platforms:
The problem is specific to the domain of quantum computing, is located in a component that is about evaluating the state of a quantum program, and manifests through incorrect output, which makes the bug non-trivial to detect.

More generally, the key findings of our study include:
\begin{itemize}
\item \percQuantumBugsAnnotated{} of all studied bugs in quantum computing platforms are quantum-specific, which motivates future work on dedicated approaches to prevent and find them.
\item Bugs are spread across various components of the studied platforms. Components that represent, compile, and optimize quantum programming abstractions are particularly prone to quantum-specific bugs.
\item While many (92 out of \totRealBugAnnotated{}) bugs manifest through program crashes, i.e., generic and easy-to-detect signs of misbehavior, quantum-specific bugs tend to more frequently cause incorrect outputs, making them harder to detect.
\item We present a hierarchy of recurring bug patterns and quantify how many bugs match each pattern. Besides patterns also known from other kinds of software, we identify ten novel, quantum-specific bug patterns, such as incorrectly ordered qubits and incorrect scheduling of low-level quantum operations.
\item Many bugs in quantum computing platforms can be fixed by changing only a few lines of code, making them an interesting target for automated program transformation and program repair techniques.
\end{itemize}

The results of this study are useful for both researchers and developers.
Researchers working on bug-related techniques can benefit from insights into this new domain and the kinds of problems it causes.
Some of our results call for quantum-specific approaches to prevent and detect kinds of bugs not targeted by traditional approaches.
Developers of quantum computing platforms can learn from our bug patterns as recurring mistakes to avoid in the future.
They can also use our results on which components are most bug-prone to guide the allocation of testing and analysis efforts.
As quantum computing platforms have clearly evolved beyond mere research prototypes, but are not yet as widely used as, e.g., traditional language implementations, our study of bugs and their characteristics has the chance to positively influence the design of future versions of these platforms.
To allow others to build upon our results, we share the set of studied bugs, including all annotations produces during the study.
For example, we envision the dataset to serve as a basis for evaluating future work on finding bugs in quantum computing platforms.

Given the young age of practical quantum computing, there currently are only few empirical studies on it.
\citet{huangQDBQuantumAlgorithms2019} describe their experience of developing and testing quantum algorithms.
A recent position paper~\cite{camposQBugsCollectionReproducible2021} underlines the need for studying quantum bugs.
Both approaches focus on developing quantum algorithms, whereas we
study bugs in quantum computing platforms.
Others have collected a dataset of 36 bugs~\cite{zhaoBugs4QBenchmarkReal2021a}, but only from a single quantum computing platform, and without a deeper analysis of the properties of these bugs.

In summary, this paper contributes the following:
\begin{itemize}
\item The first in-depth study of \totRealBugAnnotated{} real-world bugs in \totComputingPlatforms{} quantum computing platforms.
\item Insights about the components and symptoms of bugs, as well as the complexity of their fixes.
\item A hierarchy of recurring bug patterns, including ten novel, quantum-specific patterns.
\item A publicly shared dataset~\footnote{Dataset available at \url{https://doi.org/10.5281/zenodo.5834281} and \url{https://github.com/MattePalte/Bugs-Quantum-Computing-Platforms}. Note that the universally unique identifiers (UUIDs) of bugs given in the paper refer to this shared dataset.} of annotated, real-world bugs to support future work on preventing and detecting bugs in quantum computing platforms.
\end{itemize}

\section{Background}

We briefly discuss some basic quantum computing concepts (Section~\ref{sec:bg concepts}) and the typical structure of a quantum computing platform (Section~\ref{sec:bg platform}).

\subsection{Quantum Computing Concepts}
\label{sec:bg concepts}

\paragraph{Qubits}

Unlike a \textit{classical bit}, which can be either in state 0 or 1, a \textit{qubit} is a \textit{superposition} of the two states 0 and 1, written $|q\rangle = \alpha_0 |0\rangle + \alpha_1 |1\rangle$.
Upon \textit{measurement}, one can observe a probabilistic state given by the coefficients of the superposition: $|\alpha_0|^2 + |\alpha_1|^2 = 1$.
Thus, a qubit can effectively encode more information than a classical bit, which is the reason for the speedup behind quantum computing.
Once measured, a qubit collapses to state 0, with probability $|\alpha_0|^2$, or to state 1, with probability $|\alpha_1|^2$.
Multiple runs of a quantum program will return different results, as defined by these probabilities.
When encoding classical information into qubits, the order of qubits is important.
For example, programs may store the most or the least significant bit first.

\paragraph{Circuits}
Quantum programming languages provide abstractions for storing and manipulating qubits.
To store classical and quantum information, classical registers and quantum registers are available, respectively.
The most widespread model to express quantum computations is the gate model, where a program is expressed in a \textit{quantum circuit} that describes elementary operations performed on qubits in a predefined sequence.
These operations are represented with \textit{gates}, which are the building blocks of circuits, similar to classical logic gates on conventional digital circuits.
Once defined by a developer, circuits are executed either on a quantum computer or a simulator.
For mapping circuits onto hardware and for scheduling the operations, the \emph{count of qubits} used in a circuit is important.
As typically not all qubits are measured at the end, there is a distinction between total qubits and measured qubits.

\paragraph{Noise}
A quantum program is not only probabilistic due to the superposition and measurement, but it is also affected by \emph{noise} induced during a computation.
The underlying reason is the phenomenon of crosstalk~\cite{muraliSoftwareMitigationCrosstalk2020}, i.e. that the computation on some qubits physically disturbs the information stored in some neighboring qubits, together with the physical errors of executing operations on hardware.
Because of noise, statistical tests on random variables are often used to interpret results.

\subsection{Quantum Computing Platforms}
\label{sec:bg platform}

\begin{figure}[t]
  \includegraphics[width=\textwidth]{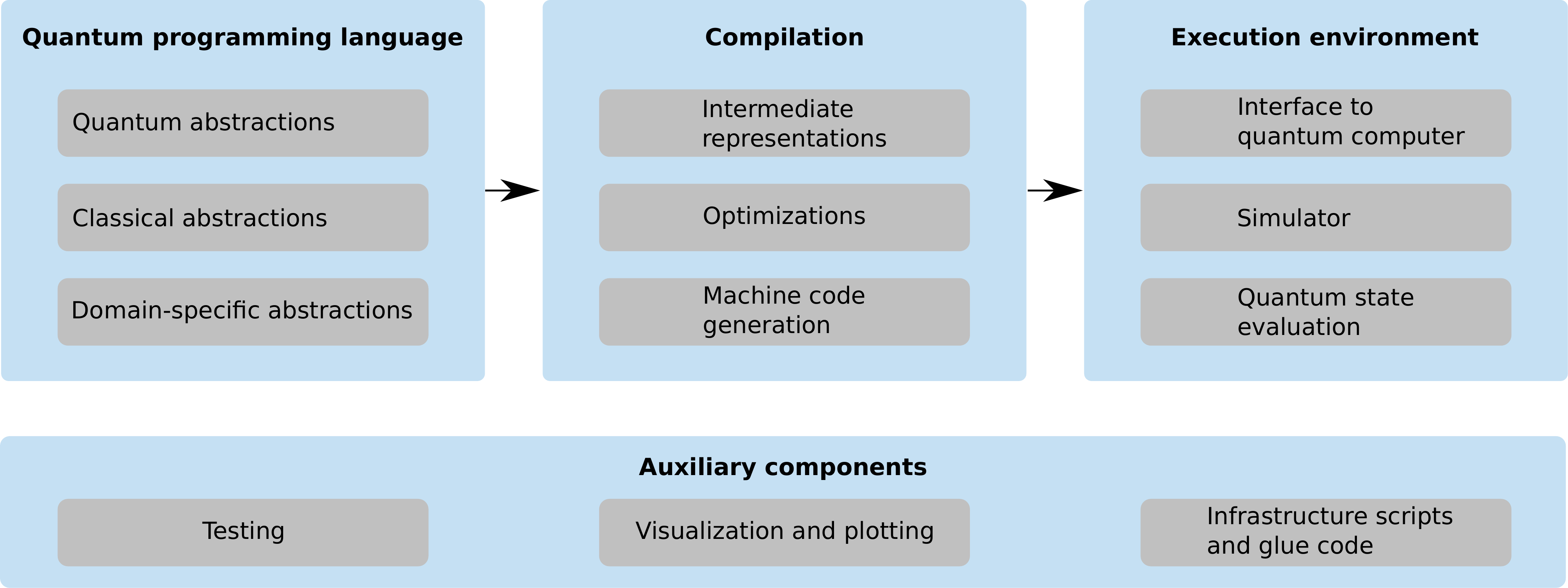}
  \centering
  \caption{Main components of quantum computing platforms.}
  \label{fig:platform_components}
\end{figure}

To express and execute quantum programs, developers build upon \emph{quantum computing platforms}, by which we mean the entire software stack that enables quantum computing.
Broadly speaking, such platforms consist of three main parts, which we further decompose into several components, as illustrated in Figure~\ref{fig:platform_components}.
First, each platform has a \emph{quantum programming language} that allows developers to express quantum algorithms.
Some ``languages'' are APIs provided as a library for a well-known host language, such as Qiskit, which provides quantum programming abstractions as a Python library.
Others are stand-alone languages, such as Q\#, Silq~\cite{DBLP:conf/pldi/BichselBGV20}, and Quipper~\cite{DBLP:conf/pldi/GreenLRSV13}.
These programming languages provide \emph{quantum abstractions}, e.g., qubits, gates, circuits, and channels, but usually also \emph{classical abstractions} that are useful to express quantum algorithms, e.g., matrices, tensors, and directed acyclic graphs.
In addition, some languages come with \emph{domain-specific abstractions}, e.g., to allow for expressing quantum algorithms in chemistry, finance, or machine learning.

Second, a quantum computing platform has a \emph{compiler} to translate programs written in the quantum programming language into a low-level instruction set.
The tasks performed in a quantum compiler resemble those known from traditional compilers.
Focusing on components most relevant for our study, Figure~\ref{fig:platform_components} shows three components of a typical quantum compiler.
The \emph{intermediate representation} includes code for creating and manipulating in-memory representations of a to-be-compiled quantum program.
Many platforms have more than one intermediate representation, e.g., a high-level, AST-like representation and an assembly-level language, such as QASM~\cite{crossOpenQuantumAssembly2017}.
To improve the efficiency of executing a quantum program, compilers implement various \emph{optimizations}, e.g., to reduce the circuit depth by aggregating, removing, or rearranging gates.
Eventually, quantum compilers have a component for \emph{machine code generation}, where ``machine'' refers to one or more execution environments, as described next.

Third, a quantum computing platform comes with support for one or more \emph{execution environments}.
Dedicated quantum computers have made impressive progress in recent years. They can be categorized into devices based on a discrete gate model, a continuous gate model, and adiabatic quantum computation~\cite{fingerhuthOpenSourceSoftware2018}.
Because quantum computers are typically operated in a computing center, developers access them through some kind of \emph{interface to a quantum computer}, similar to traditional cloud computing interfaces.
An alternative to dedicated quantum hardware are \emph{simulators}, which aim at mimicking the operations performed by a quantum computer on traditional hardware.
To faithfully simulate the execution, simulators typically include a model of quantum noise induced by stray electromagnetic fields or material defects.
An important aspect of executing a quantum program, both on dedicated hardware and in a simulator, is to evaluate the state of the program, e.g., by measuring the values of qubits after a computation.
We call the component that implements measurements \emph{quantum state evaluation}, which typically includes code for calibrating measurements and for mitigating errors.

Beyond the three main parts of quantum computing platforms, there are several auxiliary components.
With a focus on those that are most relevant for our study, Figure~\ref{fig:platform_components} shows three auxiliary components.
\emph{Testing} includes code both for testing the quantum computing platform itself and to enable developers of quantum algorithms to test their code.
To help handle the complexity of quantum programs and their results, platforms often provide a component for \emph{visualization and plotting}.
For example, this component may visualize the circuit of a quantum program or plot the probabilities that describe the possible output states.
Finally, as every software project, quantum computing platforms need \emph{infrastructure scripts and glue code}, e.g., to install the software and to connect different parts of the platform.

\section{Methodology}

This section describes how we conduct our study by formulating the research questions we address (Section~\ref{sec:RQs}), presenting the projects we study (Section~\ref{sec:projects}), describing how we identify bugs to study (Section~\ref{sec:identify bugs}), and detailing how we annotate the bugs to answer our research questions (Section~\ref{sec:annotating}).

\subsection{Research Questions}
\label{sec:RQs}

Our study is driven by five research questions:
\begin{itemize}
\item \emph{RQ1: How many of the bugs in quantum computing platforms are specific to quantum computing, as opposed to classical bugs that may also occur in other projects?}
This question is relevant to understand to what extent quantum computing platforms can benefit from traditional bug-related techniques, and whether dedicated approaches for preventing and detecting quantum bugs are needed.

\item \emph{RQ2: Where in quantum computing platforms do the bugs occur?}
Understanding which components of a platform are most bug-prone will serve as guidance on allocating efforts toward detecting and preventing bugs.
The answer to this question is relevant both for practitioners, e.g., to decide where to spend testing efforts, and for researchers, e.g., when developing novel type systems for avoiding bugs or novel bug detection techniques.

\item \emph{RQ3: How do the bugs manifest?}
This question helps understand to what degree bugs may be found through generic signs of misbehavior, e.g., a program crash, or require application-specific oracles, e.g., because a bug causes an incorrect measurement of quantum states.
In addition, knowing the consequences of bugs will make us aware of the risks incurred by leaving bugs undetected in a quantum computing platform.

\item \emph{RQ4: What recurring bug patterns do exist?}
Identifying common programming mistakes serves as a basis for creating techniques that prevent and detect specific kinds of bugs.
Furthermore, a collection of recurring bug patterns will help educate practitioners by highlighting mistakes to avoid.

\item \emph{RQ5: How complex are the bug fixes?}
Studying the complexity of patches is relevant for work toward automating the process of fixing bugs in quantum computing platforms, e.g., via synthesizing program transformations~\cite{Gao2020,Miltner2019} or automated program repair~\cite{baderGetafixLearningFix2019,cacm2019-program-repair,liDLFixContextbasedCode2020,berabiTFixLearningFix2021}.
\end{itemize}

\subsection{Selecting Projects to Study}
\label{sec:projects}

The goal of this paper is to study bugs in quantum computing platforms, where a complete platform covers all components described in Section~\ref{sec:bg platform}.
In practice, different software projects cover different parts of a complete platform.
Some projects are more focused on simulation, such as Qualcs~\cite{suzukiQulacsFastVersatile2021a}, others on interacting with quantum computing devices, such as Amazon Braket~\cite{gonzalezCloudBasedQC2021}.
Others again are specialized on domain-specific abstractions, such as Pennylane~\cite{bergholmPennyLaneAutomaticDifferentiation2020} and Tequila~\cite{kottmannTequilaPlatformRapid2021}, or on advanced error mitigation techniques, such as Mitiq~\cite{laroseMitiqSoftwarePackage2021}.
Qiskit~\cite{QiskitQiskit2021} and Cirq~\cite{developersCirq2021}, two large-scale projects backed by IBM and Google, respectively, are perhaps closest to a complete platform, but studying only them would ignore more specialized projects.

\begin{table}[h!]

\caption{Repositories considered in the study with the total commits, the commits satisfying our keyword heuristic based on ``fix'', and the sampled commits divided in real bugs and false positives. We also report the components implemented by each and the programming language: QA (quantum abstraction), CA (classical abstraction), DA (domain-specific abstraction), IR (intermediate Representation), OPT (optimization), MCG (machine code generation), INTER (interface to quantum computer), SIM (simulation), QSE (quantum state evaluation), TEST (testing), VIZ (visualization and plotting).}

\resizebox{\textwidth}{!}{%

\begin{tabular}{lrrrr|llllllllllll}
\toprule
 & \multicolumn{4}{c|}{Commits} & \multicolumn{11}{c}{Components} &      \\
 \cmidrule{2-5}  \cmidrule{6-16}
 & \multicolumn{2}{l}{} & \multicolumn{2}{l|}{Sampled} &      \\
 \cmidrule{4-5}
Repository &   Total &  Fix &     Bug &  FP &         \rot{QA} &      \rot{CA} &      \rot{DA} &      \rot{IR} &     \rot{OPT} &     \rot{MCG} &   \rot{INTER} &     \rot{SIM} &     \rot{QSE} &    \rot{TEST} &     \rot{VIZ} &     Languages \\
\midrule
\href{https://github.com/PennyLaneAI/pennylane}{ PennyLane} &    2,089 &   132 &      18 &    2 &     \ding{51} &  \ding{51} &  \ding{51} &         &         &         &         &  \ding{51} &  \ding{51} &  \ding{51} &  \ding{51} &       Py \\
\href{https://github.com/ProjectQ-Framework/ProjectQ}{ ProjectQ} &     238 &    26 &      15 &    5 &     \ding{51} &  \ding{51} &         &         &  \ding{51} &  \ding{51} &  \ding{51} &  \ding{51} &  \ding{51} &  \ding{51} &  \ding{51} &       Py \\
\href{https://github.com/QE-Lab/OpenQL}{ OpenQL} &    2,503 &    44 &      12 &    8 &     \ding{51} &  \ding{51} &         &  \ding{51} &  \ding{51} &  \ding{51} &  \ding{51} &         &         &  \ding{51} &  \ding{51} &      C++, Py \\
\href{https://github.com/Qiskit/qiskit-aer}{ Qiskit Aer} &    1,197 &   187 &      16 &    4 &     \ding{51} &  \ding{51} &         &         &  \ding{51} &  \ding{51} &         &  \ding{51} &  \ding{51} &  \ding{51} &         &      C++, Py \\
\href{https://github.com/Qiskit/qiskit-ignis}{ Qiskit Ignis} &     572 &    33 &      12 &    8 &     \ding{51} &         &  \ding{51} &         &         &  \ding{51} &         &         &  \ding{51} &  \ding{51} &         &       Py \\
\href{https://github.com/Qiskit/qiskit-terra}{ Qiskit Terra} &    5,838 &   748 &      14 &    6 &     \ding{51} &  \ding{51} &  \ding{51} &  \ding{51} &  \ding{51} &  \ding{51} &  \ding{51} &  \ding{51} &  \ding{51} &  \ding{51} &  \ding{51} &       Py \\
\href{https://github.com/aspuru-guzik-group/tequila}{ Tequila} &    1,140 &    27 &      14 &    6 &     \ding{51} &  \ding{51} &  \ding{51} &         &  \ding{51} &  \ding{51} &  \ding{51} &         &         &  \ding{51} &         &       Py \\
\href{https://github.com/aws/amazon-braket-sdk-python}{ Braket} &     396 &    53 &      11 &    9 &     \ding{51} &  \ding{51} &  \ding{51} &  \ding{51} &         &  \ding{51} &  \ding{51} &         &  \ding{51} &  \ding{51} &  \ding{51} &       Py \\
\href{https://github.com/dwavesystems/dwave-system}{ Dwawe-System} &    1,121 &    37 &       9 &   11 &     \ding{51} &  \ding{51} &         &  \ding{51} &         &  \ding{51} &         &         &  \ding{51} &  \ding{51} &         &       Py \\
\href{https://github.com/eclipse/xacc}{ XACC} &    2,370 &    58 &      18 &    2 &     \ding{51} &  \ding{51} &         &  \ding{51} &  \ding{51} &  \ding{51} &  \ding{51} &  \ding{51} &  \ding{51} &  \ding{51} &         &      C++, Py \\
\href{https://github.com/microsoft/QuantumLibraries}{ QDK Libraries} &     433 &    44 &       3 &   17 &            &  \ding{51} &  \ding{51} &         &         &         &         &         &  \ding{51} &  \ding{51} &  \ding{51} &           Q\# \\
\href{https://github.com/microsoft/qsharp-compiler}{ QDK Q\# Compiler} &     514 &    74 &      14 &    6 &            &  \ding{51} &         &  \ding{51} &  \ding{51} &  \ding{51} &  \ding{51} &         &         &  \ding{51} &         &   C\#, F\#, Q\# \\
\href{https://github.com/microsoft/qsharp-runtime}{ QDK Q\# Runtime} &     438 &    72 &      14 &    6 &     \ding{51} &  \ding{51} &         &         &         &         &         &  \ding{51} &  \ding{51} &  \ding{51} &         &  C\#, C++, Q\# \\
\href{https://github.com/quantumlib/Cirq}{ Cirq} &    2,450 &   341 &      12 &    8 &     \ding{51} &  \ding{51} &         &  \ding{51} &  \ding{51} &  \ding{51} &  \ding{51} &  \ding{51} &  \ding{51} &  \ding{51} &  \ding{51} &       Py \\
\href{https://github.com/qulacs/qulacs}{ Qualcs} &     692 &    69 &      13 &    7 &     \ding{51} &  \ding{51} &  \ding{51} &         &         &         &         &  \ding{51} &         &  \ding{51} &         &       C++, C \\
\href{https://github.com/rigetti/pyquil}{ Pyquil} &    1,101 &   109 &       9 &   11 &     \ding{51} &         &         &         &         &  \ding{51} &  \ding{51} &  \ding{51} &  \ding{51} &  \ding{51} &  \ding{51} &       Py \\
\href{https://github.com/unitaryfund/mitiq}{ mitiq} &     665 &    37 &       6 &   14 &     \ding{51} &         &         &         &         &         &  \ding{51} &         &  \ding{51} &  \ding{51} &         &       Py \\
\href{https://github.com/xanaduai/strawberryfields}{ StrawberryFields} &    1,100 &    49 &      13 &    7 &     \ding{51} &         &  \ding{51} &         &         &  \ding{51} &  \ding{51} &  \ding{51} &  \ding{51} &  \ding{51} &  \ding{51} &       Py \\
\midrule
Total &   24,857 &  2,140 &     223 &  137 &            &         &         &         &         &         &         &         &         &         &         &              \\
\bottomrule
\end{tabular}
}

\label{tab:platform_details_commits_and_components}

\end{table}

To adequately cover all aspects of quantum computing platforms we hence study not only a single or a few projects, as done in prior work~\cite{huangQDBQuantumAlgorithms2019,zhaoBugs4QBenchmarkReal2021a}, but a total of \totComputingPlatforms{} projects.
We focus on open-source projects on GitHub, enabling us to apply the same process to each project.
While this focus may potentially bias our results, we are not aware of any closed-source platform that is as mature as, e.g., Qiskit and Cirq.
To select projects, we draw inspiration from previous work~\cite{fingerhuthOpenSourceSoftware2018} and extend their list with all projects on a list of ``quantum full stack libraries'' curated by the Quantum Open Source Foundation\footnote{\url{https://qosf.org/project_list/}}.
The resulting selection covers the most popular Python libraries according to the PyPi download statistics~\cite{finkeRelativePopularityDifferent2021}, and aims at covering different components of a quantum computing platform.
Of course, the set of studied platform is not comprehensive, and other platforms, such as Tket\footnote{\url{https://github.com/CQCL/tket}}, could be added in the future.

Table~\ref{tab:platform_details_commits_and_components} shows the \totComputingPlatforms{} projects we study.
For each project, the table shows the components of a hypothetical, complete quantum platform that the project covers.
We consider a project to cover a particular component if the project repository has at least one source code file that implements a functionality of this component, which we determine by manually inspecting the code base and reading the source code.
Code that simply calls into another project to use its implementation of a component is not considered as covering the component.
Overall, Table ~\ref{tab:platform_details_commits_and_components} shows that our selection of projects covers all main components of quantum computing platforms, and that studying a single project would give only a partial view on particular components or on a single programming language.

\subsection{Identifying Bugs to Study}
\label{sec:identify bugs}

Based on the selected projects to study, we gather a set of bugs in these projects.
Following the usual definition of ``bug'', we consider a bug as a problem in the source code that causes the behavior of the program to diverge from the expected behavior.
To identify bugs in the given projects, we first automatically gather a set of commits that are likely to fix a bug, and then manually validate them to keep only actual bugs.

\subsubsection{Automated Selection of Bug Candidates}

We scan the version history of each project for commit messages that are likely to refer to bug fixes.
Similar to previous studies of bugs~\cite{karampatsisHowOftenSingleStatement2020, rayNaturalnessBuggyCode2016}, we search for the keyword ``fix'' within the first line of the commit message.
To reduce the number of coincidental matches, we further filter commits by checking that the message refers to an issue or a pull request by containing a ``\#'' followed by a number.
Ensuring that the studied bugs all refer to an issue or a pull request also helps in understanding the bugs based on the description of the issue or discussions among developers.
Finally, we ignore commits with a message that contains any of the keywords ``refactor'', ``typo'', ``requirement'', ``import'', and ``style'', as such commits often are not about bugs but other code improvements.
Table~\ref{tab:platform_details_commits_and_components} shows for each project the total number of commits and how many of them match our commit message-based filtering.
In total, there are \totBugFixCommits{} bug candidates.

\subsubsection{Manual Validation}

Given the automatically gathered bug candidates, we manually inspect a random sample of them to keep only commits that actually fix a bug.
To this end, we sample commits from the \totComputingPlatforms{} projects.
By carefully inspecting each sampled commit, we label it either as a bug or as a false positive, until having labeled 20 instances per project.
The bugs are the basis for the remainder of this paper.
By false positives we mean commits that modify the code in some other way, e.g., by fixing a comment or documentation.
If a bug-fixing commit involves other, bug-unrelated code changes, we keep it for further study and focus on the bug-related part of the commit.
For bug-fixing commits that address multiple bugs at once, we count and study each bug individually.
Table~\ref{tab:platform_details_commits_and_components} shows the number of inspect and selected commits.
Overall, we identify and study \totRealBugAnnotated{} bugs.

\subsection{Understanding and Annotating Properties of Bugs}
\label{sec:annotating}

Answering our research questions requires a solid understanding of the studied bugs.
The following describes how we inspect the bugs to annotate them with various properties.

\subsubsection{Classical vs. Quantum-Specific}

To address RQ1, we classify each bug as either classical or quantum-specific.
We consider a bug to be \emph{quantum-specific} if the mistake is in handling quantum-specific concepts, which typically implies that understanding and fixing the bug requires knowledge of the quantum programming domain.
This definition is analogous to bugs in a traditional compiler that require knowledge of the programming language's semantics to be fixed, such as when a program is miscompiled or a type soundness promise is broken.
Inversely, we consider all other bugs to be \emph{classical}, which includes mistakes that could occur outside of quantum computing platforms and that do not require quantum-specific knowledge to be fixed.
\begin{figure}[t]
\begin{lstlisting}[language=Python, numbers=none]
for field in fields:
\end{lstlisting}
\vspace{-1em}
\begin{lstlisting}[backgroundcolor=\color{pink},numbers=none]
-   if not hasattr(field, self._options):
\end{lstlisting}
\vspace{-1em}
\begin{lstlisting}[backgroundcolor=\color{celadon},numbers=none]
+   if not hasattr(self._options, field):
\end{lstlisting}
\vspace{-1em}
\begin{lstlisting}[language=Python, numbers=none]
    raise AttributeError(
        "Options field %
        "backend" %
\end{lstlisting}
\caption{Example of a classical bug where a Python API is misused. \href{https://github.com/Qiskit/qiskit-terra/commit/c00602aadf4062b65ab3862fb729a85b20bf004b}{(UUID: 733, from Qiskit Terra)}}
\label{fig:bug_733_QiskitTerra}
\end{figure}

\begin{figure}[t]
\begin{lstlisting}[language=Python, numbers=none]
def is_identity(term):
\end{lstlisting}
\vspace{-1em}
\begin{lstlisting}[backgroundcolor=\color{pink},numbers=none]
-   return len(term) == 0
\end{lstlisting}
\vspace{-1em}
\begin{lstlisting}[backgroundcolor=\color{celadon},numbers=none]
+   if isinstance(term, PauliTerm):
+       return (len(term) == 0) and (not np.isclose(term.coefficient, 0))
+   elif isinstance(term, PauliSum):
+       return (len(term.terms) == 1) and (len(term.terms[0]) == 0) and \
+           (not np.isclose(term.terms[0].coefficient, 0))
+   else:
+       raise TypeError("is_identity only checks PauliTerms and PauliSum objects!")
\end{lstlisting}
\caption{Example of a quantum bug. \href{https://github.com/rigetti/pyquil/commit/604bf264ef522f721aa2ab7ee36dd007e095f732}{(UUID: 1988, from Pyquil)}}
\label{fig:bug_1988_Pyquil}
\end{figure}

Figure~\ref{fig:bug_733_QiskitTerra} shows a classical bug, where the developer accidentally misuses the built-in Python function \code{hasattr}.
Other examples of classical bugs are missing library imports and bugs caused by confusing basic data types, such as Python's \code{tuple} and \code{list}.
In contrast, Figure \ref{fig:bug_1988_Pyquil} is an example of a quantum bug.
The bug is in the function \code{is\_identity()}, which checks if \code{term} implements an identity function on the quantum state.
The incorrect code fails to distinguish between two quantum-specific concepts, \code{PauliTerm} and \code{PauliSum}.
To fix the bug, the developer needs to understand these concepts and how they influence the identity check.
Other examples of quantum bugs are mistakes in accurately representing quantum abstractions within a compiler, or numerical computations that fail to correctly reflect the underlying quantum phenomena.

\subsubsection{Identifying Components, Symptoms, and Recurring Bug Patterns}

Addressing RQ2, RQ3, and RQ4, we annotate all bugs with respect to the component a bug is in, the symptom through which a bug manifests, and any recurring patterns a bug belongs to.
To create these annotations, we repeatedly inspect all bugs while refining the annotations and discussing unclear annotations among the authors.
While inspecting a bug, we consider not only the actual code change, but also its commit message, any associated issues, and any discussion among the developers, e.g., as part of a pull request.
We may annotate a bug with multiple components and multiple bug patterns.
For example, if the affected code is at the interface between two components, then we annotate the  bug with both components.
In contrast, each bug has a single symptom.
If we cannot associate a specific kind of annotation to a bug based on all information available to us, then we leave this bug unannotated.
In total, the annotation process results on \totElementaryAnnotations{} annotations added to the \totRealBugAnnotated{} studied bugs.

\subsubsection{Complexity of Bug Fixes}

To address RQ5, we count the number of lines of code (LoC) that are changed to fix a bug.
If the same fix is applied at multiple locations within a single commit and if each fix is independent of the others, then we count the number of lines required to fix a single location.
The rationale is that we want to estimate how difficult fixing a bug would be for an automated tool, which could be applied at multiple locations.
For commits that modify lines not directly related to fixing the bug, we count only those lines relevant for the fix.
When a commit addresses multiple bugs, then we count the affected lines for each bug separately.
In particular, we exclude any modifications of comment lines and, unless the bug is in a test, changes to test files.
The number of ``changed lines'' is the sum of added lines and removed lines, where a line that gets edited in a minor way, e.g., to replace one token with another, is counted only once.

\section{Results}

This section presents detailed answers to our five research questions.
The IDs mentioned in example bugs refer to the supplementary material, which includes the full dataset of bugs with references to their commits and all annotations created during the study.
The supplementary material will be made publicly available once the paper gets accepted.

\subsection{RQ1: Classical vs.\ Quantum-Specific Bugs}

This research question is about the ratio of classical and quantum-specific bugs in the quantum computing platforms, which helps understand whether these platforms require domain-specific approaches for handling bugs.
Based on the classification described in Section~\ref{sec:annotating}, we find that \totClassicalBugsAnnotated{} out of the \totRealBugAnnotated{} bugs are classical, whereas  \totQuantumBugsAnnotated{} are quantum-specific bugs.
That is, even though a majority of all bugs are still classical, there also is a large percentage of bugs where detecting and fixing the bug requires domain knowledge about quantum computing.

\begin{answerbox}
\textbf{Answer to RQ1}: \percQuantumBugsAnnotated{} of all studied bugs in quantum computing platforms are quantum-specific, which motivates dedicated approaches for preventing and detecting quantum bugs.
\end{answerbox}

\paragraph{Implications}
The comforting consequence of the above finding is that traditional, application-independent bug detection techniques can contribute to improving the code of quantum computing platforms.
At the same time, we see a large potential for techniques targeting quantum-specific bugs.
The results also show a need for software developers with an in-depth understanding of quantum computing.
Regarding the future relevance of this finding, the consistent portion of quantum-specific bugs (\percQuantumBugsAnnotated{}) motivates and gives additional evidence for the continuation of quantum software engineering research~\cite{zhaoQuantumSoftwareEngineering2021}.

\subsection{RQ2: Where in Quantum Computing Platforms Do the Bugs Occur?}

\begin{figure}[t]
  \includegraphics[width=\textwidth]{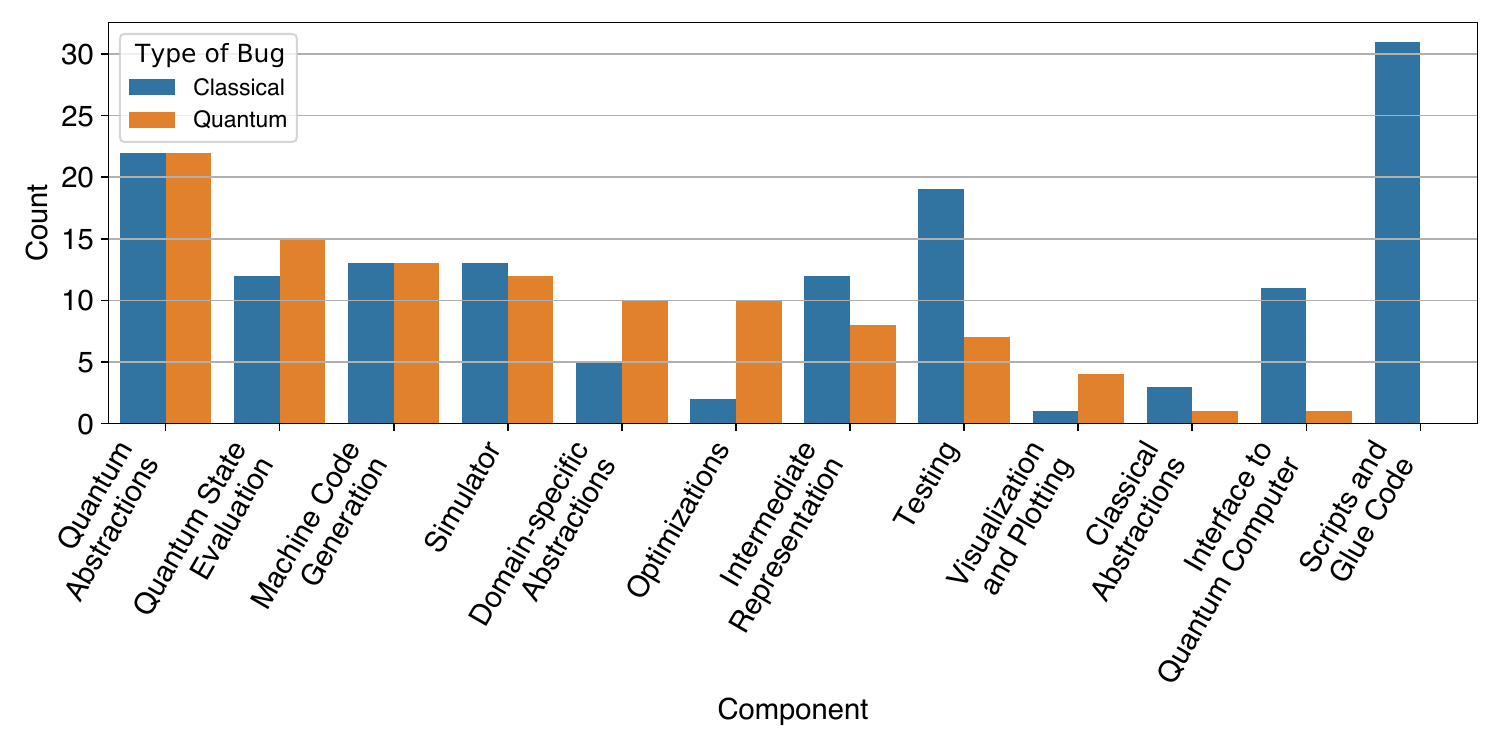}
  \centering
  \caption{Number of quantum and classical bugs per component.}
  \label{fig:quantum_vs_classical_per_component}
\end{figure}

Understanding where in quantum programming platforms bugs occur will help allocate efforts for preventing and detecting bugs to components where bugs are most likely to appear.
Figure~\ref{fig:quantum_vs_classical_per_component} shows how the studied bugs distribute across the components of a quantum computing platform.
The components are sorted in decreasing order of quantum-specific bugs per component.

We find five components where at least half of all bugs are quantum-specific: quantum state evaluation, machine code generation, domain abstractions, optimizations, and visualization and plotting.
Among those, the most striking is the optimizations component, which has almost exclusively quantum-specific bugs.
Some other components, such as quantum abstractions and simulator, show a high number of both quantum-specific and classical bugs.
For quantum abstractions, this observation can be explained by the pervasiveness of code that represents quantum primitives, such as qubits and gates, which causes a high number of mistakes.
The code of simulators usually involves the encoding of quantum operations, typically represented with matrices, and classical code to handle those large matrices, both of which can be a source of bugs.
In contrast to the above, there also are components with many classical but few or even no quantum-specific bugs, such as scripts and glue code, as well as testing.

\begin{answerbox}
\textbf{Answer to RQ2:}
Bugs occur across a wide range of components in the studied platforms.
Quantum-specific bugs are particularly prevalent in components that represent, compile, and optimize quantum programming abstractions, whereas infrastructural scripts, glue code, and testing code are mostly plagued by traditional bugs.
\end{answerbox}

\paragraph{Implications}

Components with a high number of quantum-specific bugs are most likely to benefit from quantum-specific techniques.
For example, there is a huge potential for techniques to detect bugs in optimization code, analogous to related work on analyzing traditional compilers~\cite{vafeiadisCommonCompilerOptimisations2015,baranyFindingMissedCompiler2018}.
Recent work on testing and verifying compilation and optimization passes of quantum computing platforms~\cite{shiCertiQMostlyautomatedVerification2020,hietalaVerifiedOptimizerQuantum2021,wangQDiffDifferentialTesting2021} try to exploit this potential, and our results provide an empirical justification of their assumptions.
The surprisingly large number of bugs in infrastructural scripts and glue code shows a need for better language support to prevent bugs in this component, more effective integration testing, and strong classical software engineering skills, even among quantum developers, not all of whom are computer scientists by training.
The low number of quantum-specific bugs in the interface to the hardware is a peculiarity of bugs in quantum computing, which reflects the current maturity of hardware research.
Thus this is a relevant point to monitor along with the evolution of quantum computing platforms with future empirical studies.

\subsection{RQ3: How Do the Bugs Manifest?}

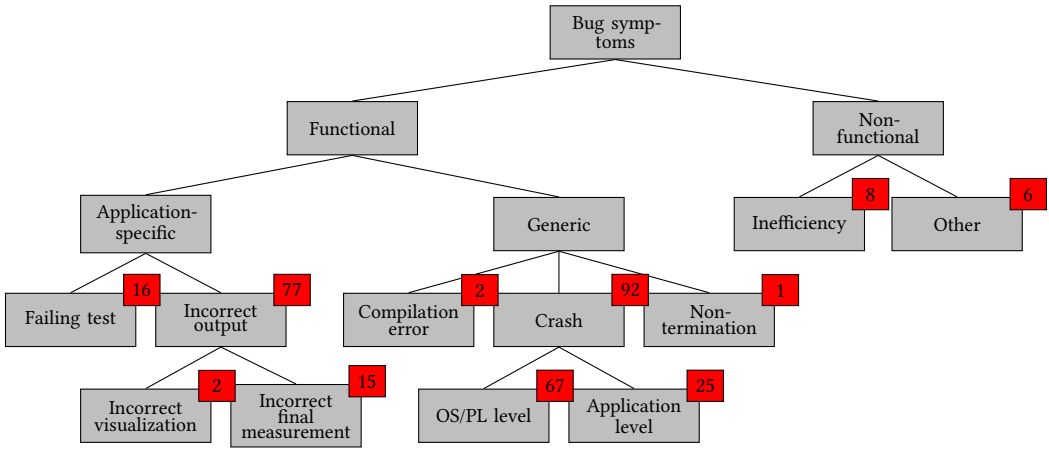
\begin{figure*}
\begin{tikzpicture}[
  level distance=1.8cm,
  every node/.style = {
    font=\fontsize{7.5}{4.5}\selectfont,
    shape=rectangle,
    draw, align=center,
    fill=lightgray,
    text width=1.5cm, minimum height=.7cm
    },
  yscale=0.7
  ]
  \tikzstyle{level 1}=[sibling distance=7cm]
  \tikzstyle{level 3}=[sibling distance=2cm]

  \node {Bug symptoms}
    child { node (funct) {Functional}[sibling distance=5.5cm]
      child { node (appl) {Application-specific}
        child { node (test) {Failing test} }
        child { node (incorOut) {Incorrect output}
          child { node (incorViz) {Incorrect visualization} }
          child { node (incorMeas) {Incorrect final measurement} }
        }
      }
      child { node (generic) {Generic}
        child { node (compil) {Compilation error} }
        child { node (crash) {Crash}
          child { node (crashOS) {OS/PL level} }
          child { node (crashApp) {Application level} }
        }
        child { node (notTerm) {Non-termination} }
      }
    }
    child { node (nonFunct) {Non-functional}[sibling distance=2.1cm]
    child { node (ineff) {Inefficiency} }
      child { node (other) {Other} }
    };

  \tikzset{every node/.style={shape=rectangle, draw, align=center, text width=.25cm, fill=red, font=\fontsize{8}{4.5}\selectfont}}
  \node[above right=-.5em and -.5em of compil] {2};
  \node[above right=-.5em and -.5em of crash] {92};
  \node[above right=-.5em and -.5em of crashApp] {25};
  \node[above right=-.5em and -.5em of crashOS] {67};
  \node[above right=-.5em and -.5em of incorMeas] {15};
  \node[above right=-.5em and -.5em of incorOut] {77};
  \node[above right=-.5em and -.5em of incorViz] {2};
  \node[above right=-.5em and -.5em of ineff] {8};
  \node[above right=-.5em and -.5em of notTerm] {1};
  \node[above right=-.5em and -.5em of other] {6};
  \node[above right=-.5em and -.5em of test] {16};
\end{tikzpicture}
\caption{Common symptoms of bugs in the quantum platforms.}
\label{fig:hierarchy_symptoms}
\end{figure*}

Understanding how bugs in quantum computing platforms manifest is a prerequisite for determining how to effectively detect them.
Figure~\ref{fig:hierarchy_symptoms} shows a hierarchy of symptoms that describe how a bug comes to the attention of users or developers.
The hierarchy summarizes similar symptoms into more general classes.
Our annotations are for those symptoms shown with a number in the figure, where the number indicates how many of all studied \totRealBugAnnotated{} bugs manifest through this symptom.

We broadly distinguish between functional and non-functional symptoms.
The functional symptoms are further classified into application-specific and generic symptoms.
One of the most prevalent functional symptoms is that the program yields an incorrect output, which is the case for \nBugsIncOutput{} of all studied bugs.
In particular, this category includes bugs that cause a quantum program to produce an incorrect final measurement and bugs that lead to an incorrect plot or diagram.
Detecting incorrect output bugs is inherently difficult, as determining what output is expected typically is highly domain-specific.
Figure~\ref{fig:bug_output_1738_Tequila} shows an example of a  bug that manifests as an incorrect measurement: The two \code{print} statements are supposed to print the same output.

\begin{figure}[t]
\begin{lstlisting}[language=Python,numbers=none]
simulator = 'qiskit'
qc  = tq.gates.X(0)
qc += tq.gates.Z(1)
result = \
  tq.simulate(qc, backend=simulator, samples=1, read_out_qubits=[0, 1])
print(result)  # Output: +1.0000|10>
result = \
  tq.simulate(qc, backend=simulator, samples=1, read_out_qubits=[1, 0])
print(result)  # Output: +1.0000|01>
\end{lstlisting}
\caption{Minimal quantum program, copied from the corresponding issue, that exposes an incorrect measurement bug. \href{https://github.com/aspuru-guzik-group/tequila/commit/be70211e1b4cad4326f01f9b2f790643255a7aaa}{(UUID: 1738, from Tequila)}}
\label{fig:bug_output_1738_Tequila}
\end{figure}

In contrast, only \nBugsFailTest{} of the application-specific bugs are detected via failing tests.
This result suggests that quantum computing platforms could benefit from more rigorous testing, possibly supported by automatically generated test suites.
The probabilistic nature of the quantum computing results and the fact that quantum states collapse to classical values when being observed poses a major challenge~\cite{huangStatisticalAssertionsValidating2019a}.

Generic, functional symptoms include compilation errors, program crashes, and non-termination.
The by far most common among those symptoms are program crashes, which we observe in \nBugsCrash{} bugs.
We further classify crashes into those induced by the operating system or the programming language, and those induced by application-specific exceptions.
The former include, e.g., runtime memory errors and runtime type errors, whereas the latter are the result of defensive programming, e.g., when code in Qiskit raises a \code{CircuitNotValid} exception if a circuit is in an unexpected state.
The large majority of program crashes (\nBugsCrashOS{} vs.\ \nBugsCrashAPP{}) is induced by the operating system or the programming language, which shows the effectiveness of generic checks, but also suggests that additional checking in the platforms could detect additional bugs.

Non-functional bugs are relatively infrequent in our study, with only \nBugsNotFunctional{} out of all \totRealBugAnnotated{} bugs.
We further classify them into inefficiencies, i.e., the program is slower than it should be, and other symptoms, e.g., a producing a correct but suboptimal result.
Since our study is based on reported and fixed bugs, one interpretation of the low number of such bugs is that detecting non-functional problems in quantum computing platforms is difficult with existing techniques.

To put these results in context, we compare against a study of bugs in deep learning compilers~\cite{shenComprehensiveStudyDeep2021}.
Those bugs also most commonly manifest through crashes and incorrect outputs, but bugs in quantum computing platforms cause fewer crashes (\percBugsCrash{} vs.\ 59.37\%) and more incorrect outputs (\percIncOutput{} vs.\ 26.26\%) than bugs in deep learning compilers.

\vspace{2mm}
\begin{figure}[t]
  \includegraphics[width=\textwidth, keepaspectratio]{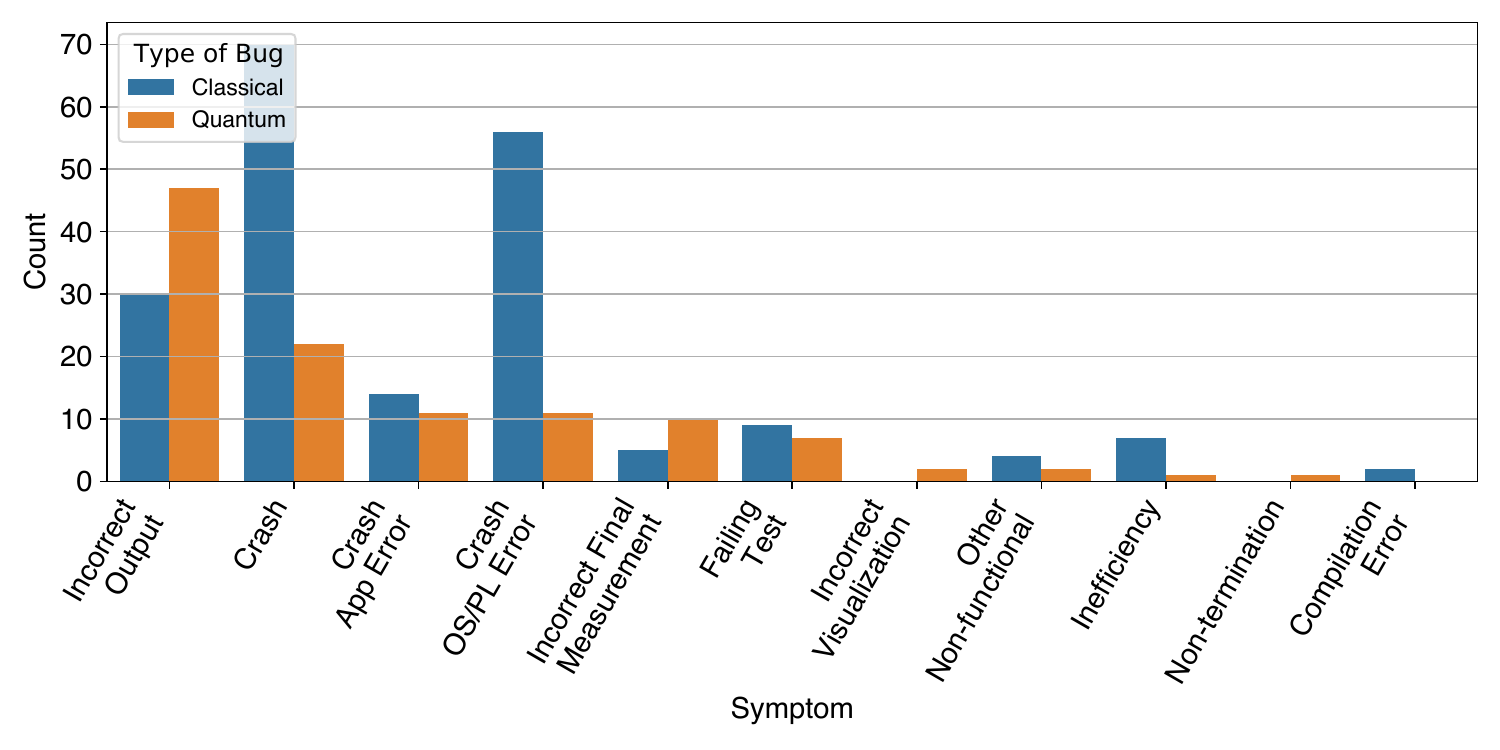}
  \centering
  \caption{Number of quantum and classical bugs per symptom.}
  \label{fig:quantum_vs_classical_per_symptom}
\end{figure}

To better understand how the different symptoms distribute over classical and quantum-specific bugs, Figure~\ref{fig:quantum_vs_classical_per_symptom} reports the number of classical and quantum bugs per symptom.
The most striking observation is that most classical bugs, but relatively few quantum bugs, manifest via crashes.
In contrast, quantum-specific bugs often manifest via symptoms that are harder to identify, such as incorrect outputs and incorrect final measurements.

\vspace{2mm}
\begin{answerbox}
\textbf{Answer to RQ3:}
Most bugs manifest through functional rather than non-functional symptoms (\nBugsFunctional{} vs. \nBugsNotFunctional{} bugs).
The two most common symptoms are crashes (\nBugsCrash{} bugs) and incorrect outputs (\nBugsIncOutput{} bugs), whereas tests are not yet the main way to find bugs in quantum computing platforms (\nBugsFailTest{} bugs).
Classical bugs in these platforms often manifest via an exception raised by the operating system or the programming language.
In contrast, quantum-specific bugs often create incorrect output, making them more difficult to detect.
\end{answerbox}

\paragraph{Implications}
The large number of bugs that manifest through application-specific symptoms call for domain-specific analysis techniques for quantum computing platforms, which could have an impact similar to the success of compiler testing over the past decade~\cite{compilerTestingSurvey2021}.
In particular, our results show a need for better support for testing quantum computing platforms.
One way to address this need are approaches for testing quantum programs, which requires specific support for the peculiar quantum characteristic~\cite{huangStatisticalAssertionsValidating2019a}.
Another promising direction is testing the platforms themselves, where we see potential for future work on differential testing and metamorphic testing, which researchers have only started to explore~\cite{wangQDiffDifferentialTesting2021}.
Finally, developers of quantum computing platforms could benefit from support for specifying their expectations about platform-internal states, e.g., in the form of invariants that intermediate representations should preserve.
Regarding the future relevance of this finding, the distribution shift of bug symptoms can be an important metric to monitor the level of maturity of a platform over time, and this study sets the first point in that sequence.
For example, crashes that now are caught at the operating system level or at the language level might be converted into more specific application-level exceptions.

\subsection{RQ4: Bug Patterns}

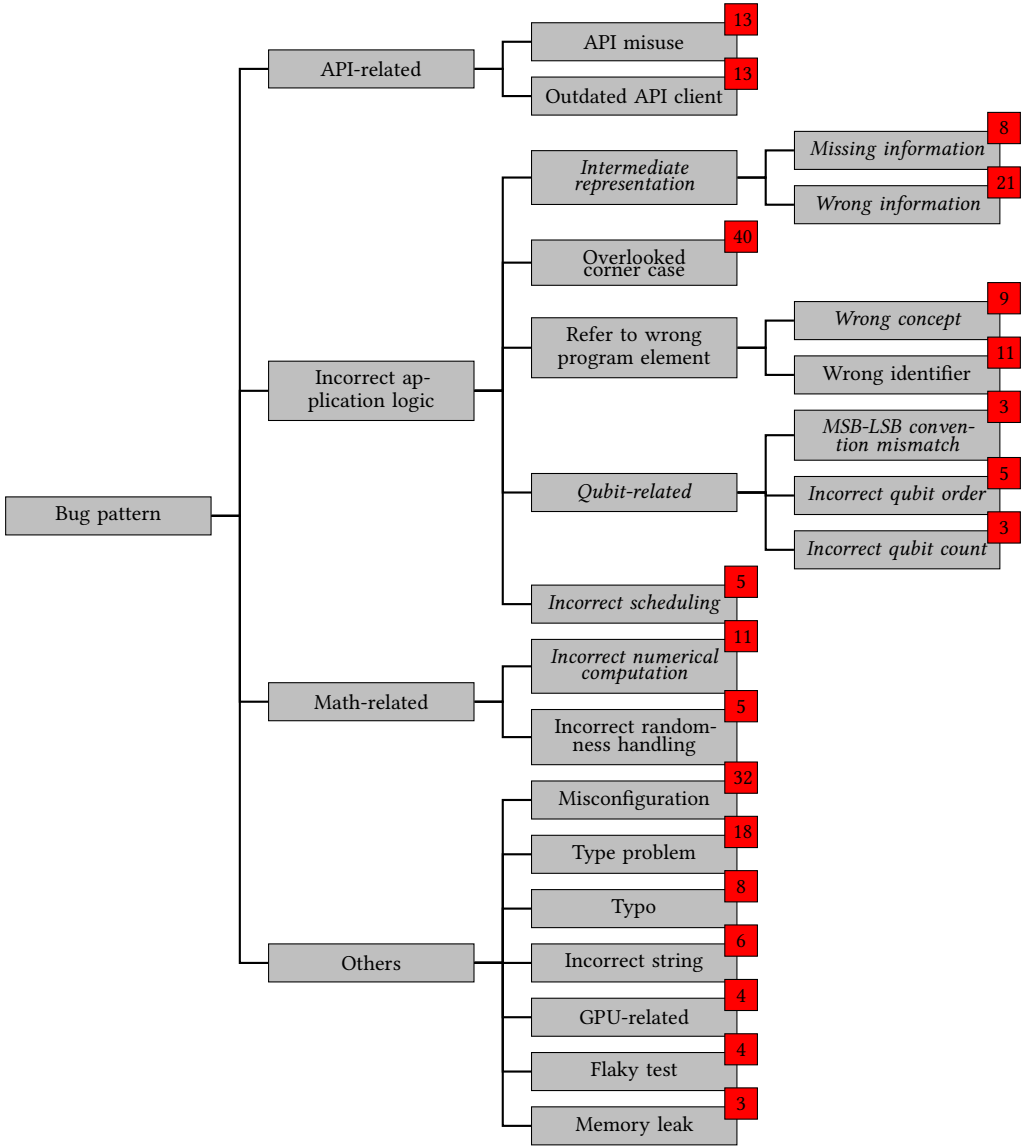
\begin{figure*}
  \begin{tikzpicture}[
    grow'=right,
    level distance=3.5cm,
    sibling distance=.2cm
  ]
  \tikzset{edge from parent/.style=
              {thick, draw, edge from parent fork right},
            every tree node/.style= {
              font=\fontsize{8}{4.5}\selectfont,
              draw,
              text width=2.5cm,
              minimum height=.5cm,
              fill=lightgray,
              align=center
            }}
  \Tree
      [.{Bug pattern}
          [.{API-related}
              [.\node(apiMisuse){API misuse}; ]
              [.\node(apiOutdated){Outdated API client}; ]
          ]
          [.{Incorrect application logic}
            [.{\emph{Intermediate representation}}
              [.\node(irMiss){\emph{Missing information}}; ]
              [.\node(irWrong){\emph{Wrong information}}; ]
            ]
            [.\node(ovCorn){Overlooked corner case}; ]
            [.{Refer to wrong program element}
              [.\node(cptSwap){\emph{Wrong concept}}; ]
              [.\node(wrongId){Wrong identifier}; ]
            ]
            [.{\emph{Qubit-related}}
              [.\node(msbLsb){\emph{MSB-LSB convention mismatch}}; ]
              [.\node(QuOrder){\emph{Incorrect qubit order}}; ]
              [.\node(QuCount){\emph{Incorrect qubit count}}; ]
            ]
            [.\node(incSched){\emph{Incorrect scheduling}}; ]
          ]
          [.{Math-related}
              [.\node(numComp){\emph{Incorrect numerical computation}}; ]
              [.\node(incRand){Incorrect randomness handling}; ]
          ]
          [.{Others}
            [.\node(misconf){Misconfiguration}; ]
            [.\node(type){Type problem}; ]
            [.\node(typo){Typo}; ]
            [.\node(string){Incorrect string}; ]
            [.\node(gpu){GPU-related}; ]
            [.\node(flakyTest){Flaky test}; ]
            [.\node(memLeak){Memory leak}; ]
          ]
      ];

      \tikzset{every node/.style={shape=rectangle, draw, align=center, text width=.2cm, fill=red, font=\fontsize{8}{4.5}\selectfont}}

  \node[above right=-.5em and -.5em of QuCount] {3};
  \node[above right=-.5em and -.5em of QuOrder] {5};
  \node[above right=-.5em and -.5em of apiMisuse] {13};
  \node[above right=-.5em and -.5em of apiOutdated] {13};
  \node[above right=-.5em and -.5em of cptSwap] {9};
  \node[above right=-.5em and -.5em of flakyTest] {4};
  \node[above right=-.5em and -.5em of gpu] {4};
  \node[above right=-.5em and -.5em of incRand] {5};
  \node[above right=-.5em and -.5em of incSched] {5};
  \node[above right=-.5em and -.5em of irMiss] {8};
  \node[above right=-.5em and -.5em of irWrong] {21};
  \node[above right=-.5em and -.5em of memLeak] {3};
  \node[above right=-.5em and -.5em of misconf] {32};
  \node[above right=-.5em and -.5em of msbLsb] {3};
  \node[above right=-.5em and -.5em of numComp] {11};
  \node[above right=-.5em and -.5em of ovCorn] {40};
  \node[above right=-.5em and -.5em of string] {6};
  \node[above right=-.5em and -.5em of type] {18};
  \node[above right=-.5em and -.5em of typo] {8};
  \node[above right=-.5em and -.5em of wrongId] {11};

    \end{tikzpicture}
    \caption{Recurring bug patterns. Quantum-specific patterns are printed in italics.}
    \label{fig:hierarchy_bug_patterns}
  \end{figure*}

The perhaps most interesting outcome of this study is a hierarchy of recurring bug patterns that we identify in the \totRealBugAnnotated{} bugs.
The bug patterns include kinds of bugs known from other domains, but also novel patterns that are specific to quantum computing platforms.
Figure~\ref{fig:hierarchy_bug_patterns} shows the hierarchy of bug patterns, along with the frequency of each pattern in our dataset.
Patterns that are specific to the domain of quantum computing platforms are printed in italics.
We identify three broad categories of bug patterns, \emph{API-related} bugs, \emph{incorrect application logic}, and \emph{math-related} bugs, as well as a set of seven patterns summarized under \emph{other}.
The following discusses each of these broad categories in detail and illustrates the patterns with examples.

\subsubsection{API-Related Bugs}

A class of bugs commonly observed in other kinds of software~\cite{zhongEmpiricalStudyAPI2020,selakovicPerformanceIssuesOptimizations2016a} are API-related bugs.
We also encounter such bugs in quantum computing platforms, and find two recurring subpatterns.
On the one hand are \emph{API misuses}, such as passing the wrong parameters, passing the right parameters in the wrong order~\cite{riceDetectingArgumentSelection2017}, or simply calling the wrong API, which account for 13 of the studied bugs.
Figure~\ref{fig:bug_1019_QiskitAer} shows an example of an API misuse bug, where the code was accidentally calling an API function not presented in the current namespace.
On the other hand are bugs due to API client code that has not yet been adapted to an API change, which we call \emph{outdated API client}, comprising a total of 13 bugs.

\begin{figure}[t]
\begin{lstlisting}[language=C, numbers=none]
    case Operations::OpType::diagonal_matrix:
\end{lstlisting}
\vspace{-1em}
\begin{lstlisting}[backgroundcolor=\color{pink},numbers=none]
-   BaseState::qreg_.apply_diagonal_matrix(op.qubits, op.params);
\end{lstlisting}
\vspace{-1em}
\begin{lstlisting}[backgroundcolor=\color{celadon},numbers=none]
+   BaseState::qreg_.apply_diagonal_unitary_matrix(op.qubits, op.params);
\end{lstlisting}
\caption{Example of API misuse on a function not present in the current namespace. \href{https://github.com/Qiskit/qiskit-aer/commit/f3b4303f478ff84bf8da15761827287d3a2f16e4}{(UUID: 1019, from Qiskit Aer)}}
\label{fig:bug_1019_QiskitAer}
\end{figure}

To further guide efforts toward finding API-related bugs, we study what kinds of APIs are incorrectly used.
Specifically, we classify all API-related bugs depending on whether the incorrect code uses a project-internal API or an API in a third-party library, referred to as ``external''.
Table \ref{tab:api_internal_external} reports the result of this classification.
We find that external APIs often are the source of outdated API client bugs, whereas the internal APIs are more commonly misused.
These results confirm the previous observation that programming against evolving third-party APIs may cause mistakes~\cite{DBLP:conf/icsm/McDonnellRK13}.

\begin{table}[t]
  \centering
  \caption{API-related bug patterns and what kinds of API they concern.}
  \begin{tabular}{lrr}
  \toprule
  Bug pattern & \multicolumn{2}{c}{Kind of API} \\
  \cmidrule{2-3}
  & Internal &  External \\
  \midrule
      API misuse &        10 &         3 \\
    Outdated API client &         5 &         8 \\
  \bottomrule
  \end{tabular}
  \label{tab:api_internal_external}
\end{table}

\subsubsection{Incorrect Application Logic}

The next family of bug patterns is about mistakes in implementing the application logic.
Naturally, many of these bug patterns are specific to the domain of quantum computing platforms.
We identify five prevalent bug patterns, which the following discusses in detail.

\paragraph{Intermediate Representation}
A common class of bugs is about corrupting an intermediate representation of a quantum program, which may occur while creating the representation or while manipulating it.
We identify two subpatterns that are about \emph{missing information} and \emph{wrong information} in the intermediate representation, which account for eight and 21 bugs, respectively.
Figure~\ref{fig:bug_136_Pennylane} gives an example of a wrong information bug, where the code fails to add some attributes while expanding the intermediate representation of a tape, which is a data structure to represent quantum circuits and measurement statistics.

\begin{figure}[t]
\begin{lstlisting}[language=Python, numbers=none]
def expand_tape(tape, depth=1, stop_at=None, expand_measurements=False):
    """Expand all objects in a tape to a specific depth."""
    ...
    new_tape = tape.__class__()
\end{lstlisting}
\vspace{-1em}
\begin{lstlisting}[backgroundcolor=\color{celadon},numbers=none]
+   new_tape.__bare__ = getattr(tape, "__bare__", tape.__class__)
\end{lstlisting}
\caption{Example of bug where a part of the intermediate representation is lost because the code forgot to update the attributes of a tape. \href{https://github.com/PennyLaneAI/pennylane/commit/1a48733e25caa373966a03dcc0d073420fcd5fed}{(UUID: 136, from Pennylane)}}
\label{fig:bug_136_Pennylane}
\end{figure}

\paragraph{Overlooked Corner Case}
A common kind of mistake, not only but also in quantum computing platforms, are overlooked corner cases.
The fix of such bugs typically expands the code to handle a rare input, which the developer forgot to consider.
With a total of 40 bugs, overlooked corner cases is the most prevalent of all bug patterns.
Figure~\ref{fig:bug_505_Qiskit_Terra} shows an example bug, where the code to compute the depth of a circuit did not consider the corner case of not having any registers.

\begin{figure}[t]
\begin{lstlisting}[language=Python, numbers=none]
def depth(self):
    """Return circuit depth (i.e., length of critical path)."""
    ...
\end{lstlisting}
\vspace{-1em}
\begin{lstlisting}[backgroundcolor=\color{celadon},numbers=none]
+   # If no registers return 0
+   if reg_offset == 0:
+       return 0
\end{lstlisting}
\caption{Example of bug where a corner case has not been considered. \href{https://github.com/Qiskit/qiskit-terra/commit/103f6ffcee69b924417f59c0cc830815b6728387}{(UUID: 505, form Qiskit Terra)}}
\label{fig:bug_505_Qiskit_Terra}
\end{figure}

\paragraph{Refer to Wrong Program Element}
This bug pattern occurs when a developer confuses two program elements, and hence, the code refers to one program element instead of another.
We distinguish two subpatterns.
The \emph{wrong concept} pattern means that the developer confuses two related concepts in the application domain of quantum computing platforms.
For example, Figure~\ref{fig:bug_1869_XACC} shows a bug where the code confuses the index of a qubits with the index of its corresponding bit.
The bug is fixed by looking up the bit via \code{qubitToClassicalBitIndex}.
Wrong concept bugs are quantum-specific, and we find a total of nine of them.

\begin{figure}[t]
\begin{lstlisting}[language=C, numbers=none]
void visit(ConditionalFunction& c) {
    auto visitor = std::make_shared<QuilVisitor>();
\end{lstlisting}
\vspace{-1em}
\begin{lstlisting}[backgroundcolor=\color{pink},numbers=none]
-   quilStr += "JUMP-UNLESS @" + c.getName() + " [" + std::to_string(c.getConditionalQubit()) + "]\n";
\end{lstlisting}
\vspace{-1em}
\begin{lstlisting}[backgroundcolor=\color{celadon},numbers=none]
+   auto classicalBitIdx = qubitToClassicalBitIndex[c.getConditionalQubit()];
+   quilStr += "JUMP-UNLESS @" + c.getName() + " [" + std::to_string(classicalBitIdx) + "]\n";
\end{lstlisting}
\vspace{-1em}
\begin{lstlisting}[language=C, numbers=none]
    for (auto inst : c.getInstructions()) {
        inst->accept(visitor);
    }
\end{lstlisting}
\caption{Example of confusion between tracking the index mapping of the different bit position (classical and quantum). \href{https://github.com/eclipse/xacc/commit/c334829352fc2b490c407b6a09fb767a16516bbf}{(UUID: 1869; from XACC)}}
\label{fig:bug_1869_XACC}
\end{figure}

The other subpattern are \emph{wrong identifier} bugs, which means the code is confusing two identifiers, e.g., because they have a similar name.
Figure~\ref{fig:bug_1122_ProjectQ} shows an example.
Wrong identifier bugs may be the result of copied-and-pasted code~\cite{liCPMinerFindingCopypaste2006} and have been addressed in work on finding VarMisuse bugs~\cite{allamanisLearningRepresentPrograms2018a,vasicNeuralProgramRepair2018,Dinella2020,kanadeLearningEvaluatingContextual2020,Hellendoorn2020,rabinUnderstandingNeuralCode2021a}.

\begin{figure}[t]
\begin{lstlisting}[language=python, numbers=none]
if token is None:
    token = getpass.getpass(prompt='IBM Q token > ')
if device is None:
\end{lstlisting}
\vspace{-1em}
\begin{lstlisting}[backgroundcolor=\color{pink},numbers=none]
-   token = getpass.getpass(prompt='IBM device > ')
\end{lstlisting}
\vspace{-1em}
\begin{lstlisting}[backgroundcolor=\color{celadon},numbers=none]
+   device = getpass.getpass(prompt='IBM device > ')
\end{lstlisting}
\caption{Example of wrong identifier bug. \href{https://github.com/ProjectQ-Framework/ProjectQ/commit/6ef8a0eeb06267b27bbbe9207b71013e9cca65d4}{(UUID: 1122, from ProjectQ)}}
\label{fig:bug_1122_ProjectQ}
\end{figure}

\paragraph{Qubit-Related}
As qubits are the basic unit of quantum information, it may come as no surprise that we find several qubit-related bug patterns.
The first pattern refers to a mistake in representing multiple qubits in memory, where the code stores the most significant bit (MSB) first, instead of the least significant bit (LSB), or vice versa.
We call this pattern \emph{MSB-LSB convention mismatch}.
The second bug pattern is about computing in incorrect qubit count, e.g., in a function that counts the number of qubits required by a circuit.
Finally, the third bug pattern is about code that causes incorrectly ordered qubits, e.g., when a mapping of the qubit indices is not properly maintained.
Figure~\ref{fig:bug_1900_xacc} shows a bug in an optimization pass to merge gates, which suffers from a MSB-LSB convention mismatch.
The developers fix the problem by reversing the qubits before using them.

\begin{figure}[t]
\begin{lstlisting}[language=C, numbers=none]
void GateFuser::visit(CNOT& cnot) {
    Eigen::MatrixXcd cnotMat{ Eigen::MatrixXcd::Zero(4, 4) };
    cnotMat << 1, 0, 0, 0,
                0, 1, 0, 0,
                0, 0, 0, 1,
                0, 0, 1, 0;
\end{lstlisting}
\vspace{-1em}
\begin{lstlisting}[backgroundcolor=\color{pink},numbers=none]
-   m_gates.emplace_back(cnotMat, cnot.bits());
\end{lstlisting}
\vspace{-1em}
\begin{lstlisting}[backgroundcolor=\color{celadon},numbers=none]
+   m_gates.emplace_back(cnotMat, reverseBitIdx(cnot.bits()));
\end{lstlisting}
\vspace{-1em}
\begin{lstlisting}[numbers=none]
}
\end{lstlisting}
\caption{Example of a quantum-specific bug caused by confusing most significant bit with least significant bit representation. \href{https://github.com/eclipse/xacc/commit/771e4f8ab46e7b29204d87541da34b76b081e855}{(UUID: 1900, from XACC)}.}
\label{fig:bug_1900_xacc}
\end{figure}

\paragraph{Incorrect Scheduling}
Our final bug pattern among the application logic-related patterns is about code that schedules the low-level instructions to be executed on a quantum computation device.
If such code accidentally schedules an instruction to be executed at the wrong timestep, we call it an instance of the \emph{incorrect scheduling} bug pattern.
We find five bugs that match this pattern, all of which are in the machine code generation component.
Figure~\ref{fig:bug_1245_OpenQL} shows an example, where the fix adapts the incorrect starting time of instructions scheduled in OpenQL.

\begin{figure}[t]
\begin{lstlisting}[language=c, numbers=none]
void PrintCCLighQasm(Bundles & bundles, bool verbose=false) {
    ...
\end{lstlisting}
\vspace{-1em}
\begin{lstlisting}[backgroundcolor=\color{pink},numbers=none]
-   size_t curr_cycle=1;
\end{lstlisting}
\vspace{-1em}
\begin{lstlisting}[backgroundcolor=\color{celadon},numbers=none]
+   size_t curr_cycle=0; //first instruction should be with pre-interval 1, 'bs 1'
\end{lstlisting}
\caption{Example of an incorrect scheduling bug. \href{https://github.com/QE-Lab/OpenQL/commit/bc0d1c7981a71349d6b1fd816276df8645dfc010}{(UUID: 1245, from OpenQL)}}
\label{fig:bug_1245_OpenQL}
\end{figure}

\subsubsection{Math-Related Bugs}

Some parts of quantum computation platforms implement mathematically modeled phenomena, which poses a risk of introducing math-related bugs.
We find 16 such bugs in our study, and classify them into two subpatterns.
The first pattern are \emph{incorrect numerical computations}, i.e., code that uses a wrong mathematical formula or model to represent a phenomenon.
For example, Figure~\ref{fig:bug_1496_Cirq} shows a bug in the implementation of the \code{\_\_rpow\_\_} operation on the \code{PauliString}, which was incorrectly dividing by four instead of two.
Understanding this and other incorrect numerical computations requires a solid understanding of the mathematics behind quantum computing.

\begin{figure}[t]
\begin{lstlisting}[language=python, numbers=none]
return pauli_string_phasor.PauliStringPhasor(
  PauliString(qubit_pauli_map=self._qubit_pauli_map),
\end{lstlisting}
\vspace{-1em}
\begin{lstlisting}[backgroundcolor=\color{pink},numbers=none]
-   exponent_neg=+half_turns / 4,
-   exponent_pos=-half_turns / 4)
\end{lstlisting}
\vspace{-1em}
\begin{lstlisting}[backgroundcolor=\color{celadon},numbers=none]
+   exponent_neg=+half_turns / 2,
+   exponent_pos=-half_turns / 2)
\end{lstlisting}
\caption{Example of a quantum bug due to an incorrect numerical computation. \href{https://github.com/quantumlib/Cirq/commit/df52d1c6168d2f64cef4045b1c4a149c98e807d3}{(UUID: 1496, from Cirq)}}
\label{fig:bug_1496_Cirq}
\end{figure}

The second math-related bug pattern is \emph{incorrect randomness handling}, which means that code related to probabilities and randomness uses these concepts incorrectly.
Figure~\ref{fig:bug_27_Qiskit_Ignis} shows an example. The bug is due to missing initialization of a random seed, which causes all subsystems of the same size within a randomized benchmark to have exactly the same gate instructions.

\begin{figure}[t]
\begin{lstlisting}[language=python, numbers=none]
def randomized_benchmarking_seq(nseeds: int = 1, ...):
    ...
        for _ in range(length_multiplier[rb_pattern_index]):
\end{lstlisting}
\vspace{-1em}
\begin{lstlisting}[backgroundcolor=\color{celadon},numbers=none]
+           # make the seed unique for each element
+           if rand_seed:
+               rand_seed += (seed + 1)
\end{lstlisting}
\vspace{-1em}
\begin{lstlisting}[language=python, numbers=none]
            new_elmnt = rb_group.random(rb_q_num, rand_seed)
\end{lstlisting}
\caption{Example of incorrect randomness handling. \href{https://github.com/Qiskit/qiskit-ignis/commit/0af27dd1fdd5c4e4d2ef80d21d0375d11a1625fd}{(UUID: 27, from Qiskit-Ignis)}}
\label{fig:bug_27_Qiskit_Ignis}
\end{figure}

\subsubsection{Other Recurrent Bug Patterns}

\begin{figure}[t]
\begin{lstlisting}[language=Python, numbers=none]
def beamsplitter(self, t, r, mode1, mode2):
    """Perform a beamsplitter operation on the specified modes.
\end{lstlisting}
\vspace{-1em}
\begin{lstlisting}[backgroundcolor=\color{pink},numbers=none]
-   t (complex): transmittivity parameter
\end{lstlisting}
\vspace{-1em}
\begin{lstlisting}[backgroundcolor=\color{celadon},numbers=none]
+   t (float): transmittivity parameters
\end{lstlisting}
\vspace{-1em}
\begin{lstlisting}[language=Python, numbers=none]
    ...
\end{lstlisting}
\vspace{-1em}
\begin{lstlisting}[backgroundcolor=\color{celadon},numbers=none]
+   if isinstance(t, complex):
+       raise ValueError("Beamsplitter transmittivity t must be a float.")
\end{lstlisting}
\caption{Example of type problem. \href{https://github.com/xanaduai/strawberryfields/commit/dd7668c30c0e20730ea1d80cab91c7247e71af5f}{(UUID: 1820, from Strawberry Fields)}}
\label{fig:bug_1820_SB_Field}
\end{figure}

Beyond the three families of bug patterns described above, we find seven additional patterns, shown under \emph{others} in Figure~\ref{fig:hierarchy_bug_patterns}.
These seven patterns are not quantum-specific, and hence described only briefly.
With \nBugsMisconfiguration{} examples, the largest category are \emph{misconfiguration} bugs, which means that an incorrect configuration parameter causes the build scripts, testing scripts, or installation scripts to perform in an unexpected way.
Another frequent pattern are \emph{type problems}, such as the example in Figure~\ref{fig:bug_1820_SB_Field}, where the bug fix raises an error if an incorrect type is passed as a parameter.
We also observe multiple examples of simple \emph{typos} in the code, \emph{string-related} bugs~\cite{eghbaliNoStringsAttached2020}, \emph{flaky tests}, \emph{memory leaks}, and \emph{GPU-related} bugs.

\subsubsection{Distribution of Classical vs.\ Quantum-Specific Bugs Across the Bug Patterns}

\vspace{2mm}
\begin{figure}[t]
  \includegraphics[width=\textwidth, keepaspectratio]{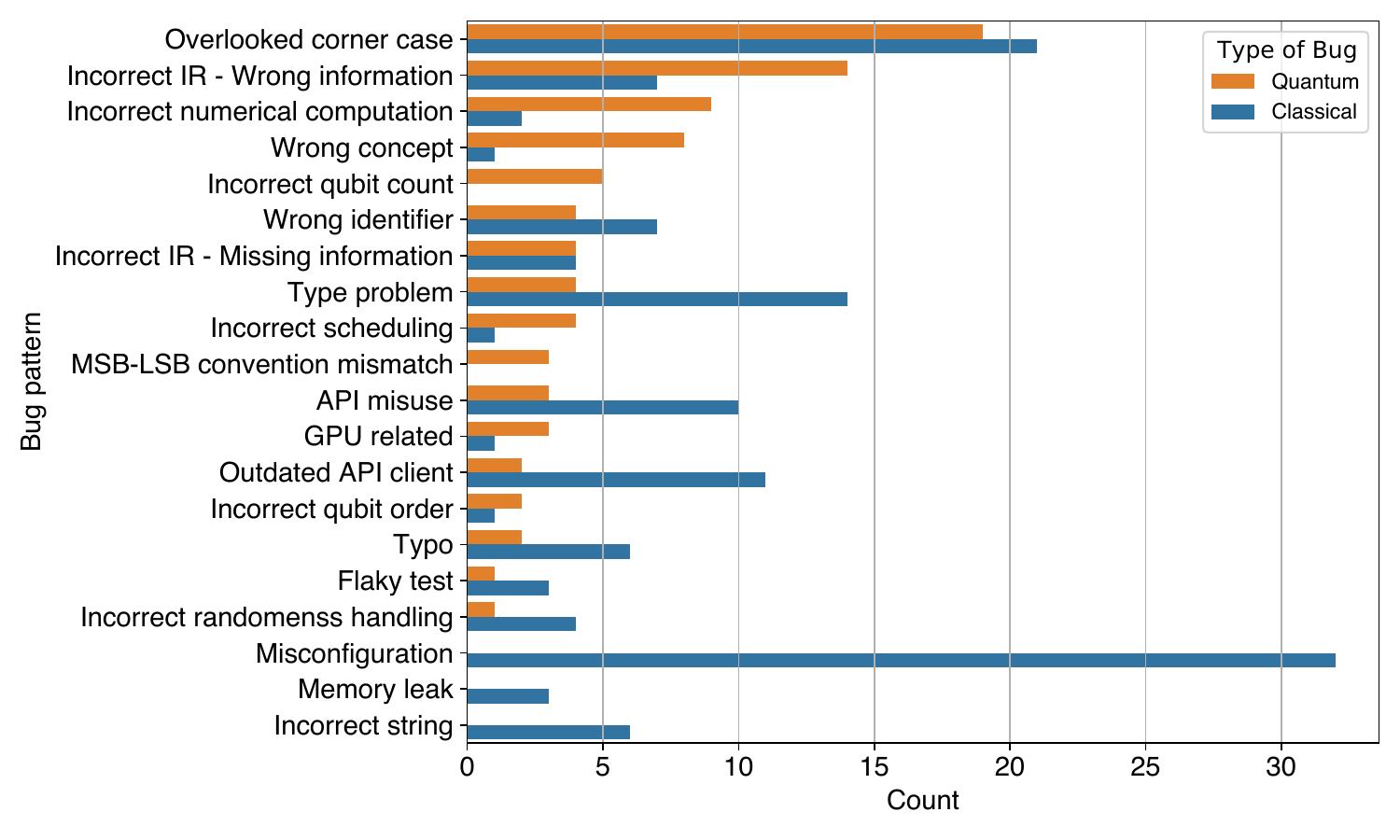}
  \centering
  \caption{Number of quantum-specific and classical bugs per bug pattern.}
  \label{fig:quantum_vs_classical_per_bug_pattern}
\end{figure}

The bug patterns described above and our classification of each bug into quantum-specific or classical (RQ1) are, a priori, independent of each other.
The following studies whether some bug patterns are particularly prevalent among quantum-specific or classical bugs.
Figure~\ref{fig:quantum_vs_classical_per_bug_pattern} shows for each bug pattern the number of quantum-specific and classical bugs.

The results allow for several interesting observations.
First, some bug patterns are clearly dominated by quantum-specific bugs.
This is obviously the cases for our novel, quantum-specific bug patterns, such as the qubit-related patterns, but also for incorrect numerical computations and wrong information in the intermediate representation.
Second, wrong concept bugs are more likely to be quantum-specific than classical, suggesting that concepts in quantum programming, such as qubits and circuits, are likely causes of confusion among developers.
Third, some bug patterns are populated both by many quantum-specific bugs and by many classical bugs, such as overlooked corner cases.
Finally, bug patterns known from other application domains, e.g., misconfigurations and type problems, are mostly found among the classical bugs.

\vspace{2mm}
\begin{answerbox}
\textbf{Answer to RQ4:}
Among the \totRealBugAnnotated{} studied bugs, there are various recurring bug patterns.
We find both patterns known from other domains, e.g., incorrect API uses (26 bugs) and type problems (18 bugs), and quantum-specific bug patterns, e.g., wrong or incomplete intermediate representations (29 bugs), mistakes related to the order or count of qubits (8 bugs), and mistakes in encoding the underlying mathematical formulas into numeric computations (11 bugs).
\end{answerbox}

\paragraph{Implications}

Our analysis of recurring bug patterns motivates several lines of future work.
First, the fact that there are quantum-specific bug patterns shows a need for new techniques that target such bugs.
For example, we envision bug detection tools that address common yet domain-specific problems in a pattern-by-pattern basis, similar to existing static bug detectors~\cite{aftandilianBuildingUsefulProgram2012}.
Our annotated bug dataset can serve as a starting point for measuring the detection abilities of such bug detectors.
Second, the observation that there are almost no bugs that are both type-related and quantum-specific motivates work on type systems to reason about quantum abstractions.
For example, such type systems could help reduce wrong concept bugs by representing different concepts with different types.
Third, the prevalence of incorrect numerical computation bugs underlines the need for developers of quantum computing platforms to carefully check the implementations of the underlying mathematical concepts.
Fourth, generic end-to-end approaches, such as differential testing, could be adapted to find bugs in quantum computing platforms, as shown by \citet{wangQDiffDifferentialTesting2021}. We also expect existing techniques for generating realistic programs, such as based on learning from a corpus of existing programs, to be useful for detecting quantum-related bugs.
Finally, most quantum-specific bugs, such as incorrect qubit order, incorrect scheduling, or incorrect numerical computation, happen in components independent of the execution environment.
This fact implies that, while waiting for future error-corrected quantum computers, the use of simulators is a good and viable option to test quantum platforms since execution on simulators can successfully trigger a large part of these bug patterns.

\subsection{RQ5: How Complex Are the Bug Fixes?}

\begin{figure}[t]
  \includegraphics[width=\textwidth]{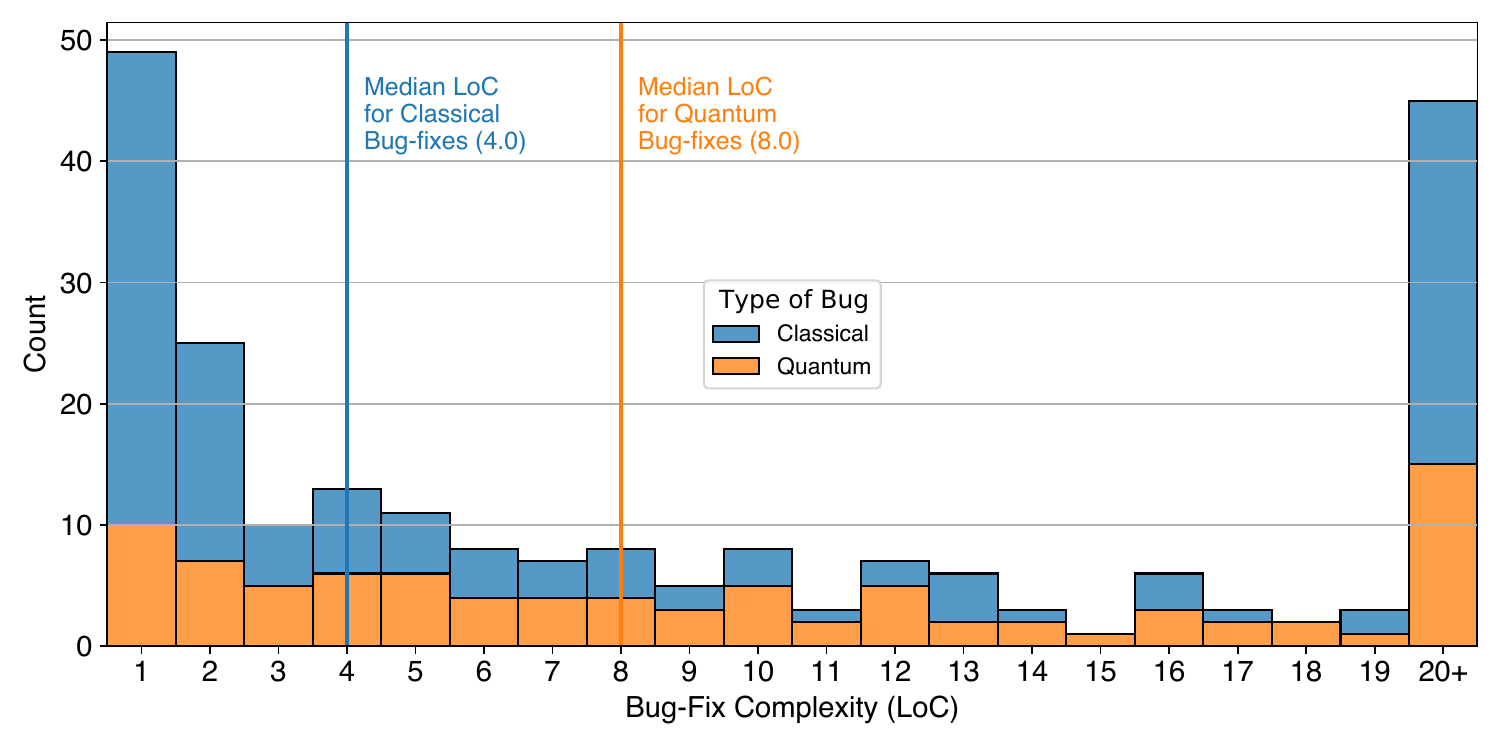}
  \centering
  \caption{Distribution of number of lines of code (LoC) for the bugs under study, divided by quantum and classical bugs.}
  \label{fig:quantum_vs_classical_and_complexity}
\end{figure}

Motivated by the recent progress on synthesizing program transformations~\cite{Gao2020,Miltner2019} and automated program repair~\cite{baderGetafixLearningFix2019,cacm2019-program-repair,liDLFixContextbasedCode2020,berabiTFixLearningFix2021}, we study the complexity of the bug fixes applied in quantum computing platforms.
We measure complexity as the number of lines changed to fix a bug.
Figure~\ref{fig:quantum_vs_classical_and_complexity} shows the number of bugs with a specific number of changed lines in its fix.
The stacked bars show the number of quantum-specific and classical bugs.
For example, there are \nBugsQuantumOneLiner{} quantum-specific bugs and \nBugsClassicalOneLiner{} classical bugs that can be fixed by changing only one line.

\begin{figure}[t]
  \centering
  \begin{minipage}{.48\textwidth}
    \centering
    \includegraphics[width=.95\linewidth]{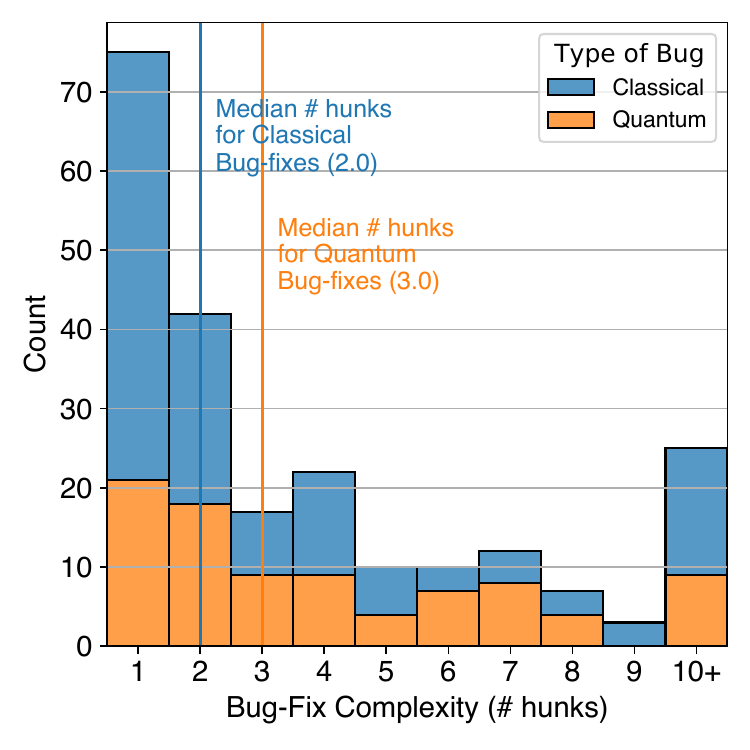}
    \captionof{figure}{Distribution of code change hunks per bug fix.}
    \label{fig:quantum_vs_classical_and_complexity_hunks}
  \end{minipage}%
  \hspace{1em}
  \begin{minipage}{.48\textwidth}
    \centering
    \includegraphics[width=.95\linewidth]{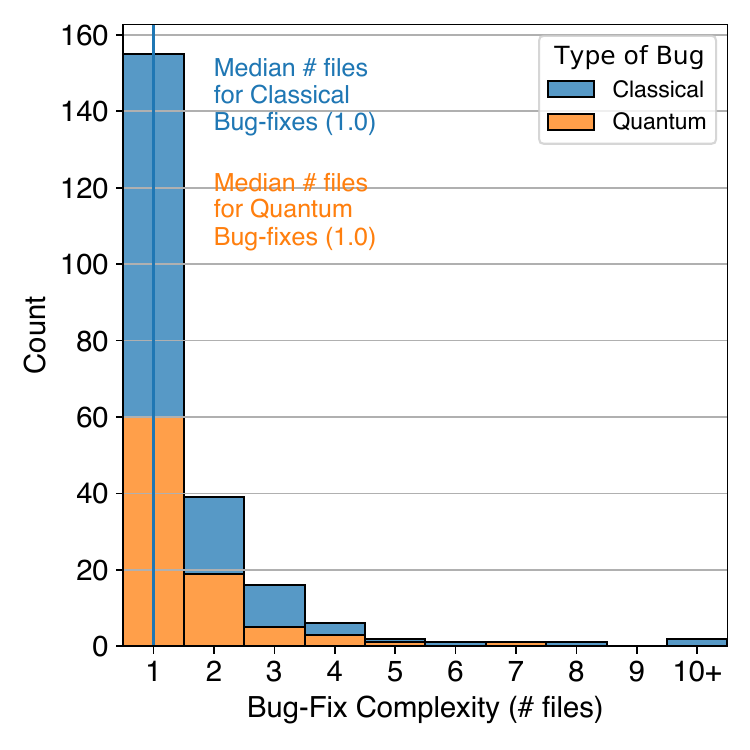}
    \captionof{figure}{Distribution of number of edited files per bug fix.}
    \label{fig:quantum_vs_classical_and_complexity_files}
  \end{minipage}
\end{figure}

Overall, we observe a similar distribution of bug fix complexity across quantum-specific and classical bugs.
Both groups of bugs include a large number of problems that can be fixed by changing only one or two lines, but also a relatively large share of bugs that need at least 20 changed lines.
Computing the median size of a bug fix for both quantum-specific and classical bugs, we obtain \medianLoCQuantum{} and \medianLoCClassical{}, respectively.
The difference suggests that despite the overall similarity of the two distributions, fixing quantum-specific bugs tends to require larger code changes.

Beside the number of changed lines, other measures of bug fix complexity exist, such as the number of methods, hunks, classes, files or packages.
We report for each bug fix also the number of code changes in Figure~\ref{fig:quantum_vs_classical_and_complexity_hunks} and changed files in Figure~\ref{fig:quantum_vs_classical_and_complexity_files}.
Note that we minimize the bug fix commit via manual inspection and leave only the relevant code change hunks that fix the bug.
For example, this might involve inspecting sub-commits of a pull request to separate the bug-fixing changes from those that refactor the code or add regression tests.
The majority of bug fixes contains less than five code changes and less than two changed files.

\begin{answerbox}
\textbf{Answer to RQ5:}
Many bugs in quantum computing platforms can be fixed by changing only one or two lines.
While this observation holds for both quantum-specific and classical bugs, the former also include a large number of bug fixes with 20 or more lines.
Most of these bugs can be fixed by editing only one or two files.
\end{answerbox}

\paragraph{Implications}
Quantum-specific bugs that can be fixed by changing only a few lines are an attractive target for automated program transformation and program repair techniques.
Yet, we also observe a disproportionally large number of bug fixes with 20 or more lines, which likely brings them out of the reach of today's bug fix automation approaches.
Regarding the impact of this finding on future work, the quantification of the scope of bug fixes in terms of the number of lines, hunks, and files can guide the design of future program repair techniques.
Moreover, regarding the area of software evolution, to the best of our knowledge, this work is the first to quantify these metrics, giving a reference to future work.

\section{Related Work}

\subsection{Studies of Quantum Bugs}

The need for and the challenges of studying quantum bugs is discussed also by a short position paper~\cite{camposQBugsCollectionReproducible2021}, which however does not address this challenge.
\citet{zhaoBugs4QBenchmarkReal2021a} provide a dataset of 36 bugs gathered from a single quantum computing platform (Qiskit).
Our study contributes both in terms of scale and depth, by studying an order of magnitude more bugs than the only existing bug dataset and by performing a detailed analysis of the bugs.
Importantly, our study is based on bugs from \totComputingPlatforms{} different projects, allowing us to draw more general conclusions than a study on a single platform can.

Related to our hierarchy of bug patterns, \citet{zhaoIdentifyingBugPatterns2021a} define eight bug patterns that focus on misuses of features of the Qiskit language.
In contrast to our work, these patterns are not based on bugs found in the wild, and hence, it remains unclear whether, and if yes, how often the bug patterns occur.
Moreover, their patterns are about quantum programs written on top of a quantum computing platform, whereas we focus on bugs in the platforms themselves.
\citet{huangQDBQuantumAlgorithms2019} study three quantum algorithms and, by implementing them in the Scaffold and ProjectQ languages, the authors propose six quantum bug patterns based on their experience while programming those algorithms.
Instead of reporting our own experience, we study real-world bugs across \totComputingPlatforms{} projects that many developers contribute to.

\subsection{Studies of Other Kinds of Bugs}

Prior work studies various other kinds of bugs.
Most closely related to our work are studies of bugs in widely used frameworks and platforms.
For example, \citet{chouEmpiricalStudyOperating2001} study bugs in operating systems, while more recent work studies bugs in compilers~\cite{DBLP:conf/issta/SunLZS16} and deep learning libraries~\cite{islamComprehensiveStudyDeep2019,shenComprehensiveStudyDeep2021}.
Our work is similar in that we study bugs in a platform used by various applications, but we cover a domain missed by prior work.
Other studies are about specific kinds of bugs, e.g., concurrency bugs~\cite{luLearningMistakesComprehensive2008a}, performance bugs~\cite{jinUnderstandingDetectingRealworld2012a,selakovicPerformanceIssuesOptimizations2016a,hanEmpiricalStudyPerformance2016}, and string-related bugs~\cite{eghbaliNoStringsAttached2020}.

\subsection{Correctness of Quantum Computing and Other Platforms}

Several approaches to increase the correctness of quantum computing platforms have been proposed recently.
\citet{shiCertiQMostlyautomatedVerification2020} describe a verification framework to check the correctness of Qiskit's compiler passes in a semi-automatic way.
\citet{wangQDiffDifferentialTesting2021} use differential testing~\cite{mckeemanDifferentialTestingSoftware1998a} to test several quantum computing platforms against each other.
The results of our study will be useful for guiding future verification and testing efforts towards components and bug patterns that are not sufficiently addressed by today's approaches.
Beyond quantum computing platforms, other widely used platforms are subject to testing and verification approaches, e.g., in the form of compiler testing~\cite{yangFindingUnderstandingBugs2011a,EMI,leFindingDeepCompiler2015,zhangSkeletalProgramEnumeration2017,compilerTestingSurvey2021}, compiler verification~\cite{leroyFormalVerificationRealistic2009}, or automated testing of deep learning libraries~\cite{DBLP:conf/icse/WangSSWWN21,CRADLE}.
Apart from the few existing approaches cited above, adapting these ideas to quantum computing platforms remains as a promising line of future work.

\subsection{Correctness of Quantum Programs}

Finding bugs in quantum programs, i.e., programs that run on a quantum computing platform, is the primary goal of another line of research.
These program are more difficult to debug than a traditional program due to the impossibility of copying quantum information~\cite{woottersSingleQuantumCannot1982} and the inherently probabilistic nature of measurements.
To tackle these challenges, \citet{huangStatisticalAssertionsValidating2019a} propose statistical methods to perform assertions in a quantum program.
\citet{Li2020b} describe a projection-based runtime assertion scheme for quantum programs that ensures that testing an assertion does not affect the tested state if it satisfies the assertion.
\citet{Yu2021} propose an abstract interpretation-based static analysis for quantum programs. %
These approaches and work on ensuring the correctness of platforms, such as ours, are complementary to each other, but share the overall goal of mitigating the risk of bugs in quantum computing.

\subsection{Quantum Programming Languages and Their Implementations}

Quantum programming languages and their implementation are an active field research.
One line of work is about language constructs to simplify particularly tricky aspects of quantum programming, such as uncomputation, which is about resetting temporary quantum values, usually before discarding them.
Convenience functions that simplify this step are proposed as \code{ApplyWith} in Q\#~\cite{svoreEnablingScalableQuantum2018} or \code{with\_computed} in
Quipper~\cite{greenQuipperScalableQuantum2013}.
\citet{DBLP:conf/pldi/ParadisBSV21} automatically synthesize uncomputation code for quantum circuits.
The bug patterns identified in our study could motivate other language constructs or code synthesis techniques.

Another line of work is about optimizing the execution of quantum programs.
\citet{DBLP:journals/pacmpl/HanerHT20} propose an optimization technique based on assertions about entanglements between qubits, which is implemented in the ProjectQ platform.
\citet{DBLP:journals/pacmpl/MeuliSRH20} describe an optimization aimed at avoiding numeric approximation errors while reducing the cost of computing with high accuracy, which is implemented in the Q\# platform.
Gleipnir computes bounds on the errors caused by noise in quantum computations, which can help in evaluating how effectively quantum compiler optimizations mitigate errors~\cite{DBLP:conf/pldi/TaoSYHCG21}.
Our study finds the optimization component to have a particularly high ratio of quantum-specific bugs, showing that correctly implementing an optimization deserves particular attention.

\section{Conclusions}

Motivated by the increasing importance of quantum computing platforms, this paper presents the first empirical study of bugs in these platforms.
Based on a set of \totRealBugAnnotated{} real-world bugs from \totComputingPlatforms{} open-source projects, we study how many bugs are quantum-specific, where the bugs occur, how they manifest, whether there are any recurring bug patterns, and how complex it is to fix the bugs.
We find that quantum-specific bugs are common and identify a novel set of quantum-specific bug patterns.
These findings show that, while platform developers can benefit from existing bug-related tools, there is a need for new, quantum-specific techniques to prevent, detect, and fix bugs.
For example, future work could design type systems to prevent developers from confusing related quantum concepts, language constructs to encode the order of qubits, static bug detectors that target quantum-specific bug patterns, and generate quantum programs to test quantum computing platforms.
Our study and its associated dataset provide concrete guidance for these research directions, and a starting point for evaluating such approaches.

\begin{acks}
This work was supported by the European Research Council (ERC, grant agreement 851895), and by the German Research Foundation within the ConcSys and Perf4JS projects.
\end{acks}

\bibliographystyle{ACM-Reference-Format}
\interlinepenalty=10000
\bibliography{phd-mattepalte.bib,referencesMichael,referencesMore}

\end{document}